\documentclass[10pt, journal, onecolumn, letterpaper, twoside]{IEEEtran}%
\usepackage{amssymb}
\usepackage{amsmath}
\usepackage{bm}
\usepackage{amsbsy}
\usepackage[mathcal]{eucal}
\usepackage{cite}
\usepackage{graphicx}
\usepackage{subfigure}
\usepackage{amsfonts}%
\begin{document}

\title{Keyhole and Reflection Effects in Network Connectivity Analysis}
\author{Mohammud Z. Bocus, \textit{Member IEEE}, Carl P. Dettmann, Justin P. Coon, \textit{Senior Member IEEE} and Mohammed R. Rahman\thanks{M. Z. Bocus is with the
Telecommunications Research Laboratory, Toshiba Research Europe Ltd., 32 Queen
Square, Bristol, BS1 4ND, U.K.; e-mail: zubeir.bocus@toshiba-trel.com.}
\thanks{C. P. Dettmann and M. R. Rahman are with the University of Bristol School of
Mathematics, University Walk, Bristol, BS8 1TW, U.K.; email: carl.dettmann@bristol.ac.uk, mamrr@bristol.ac.uk.}
\thanks{J. P. Coon is with the Department of
Engineering Science, University of Oxford, OX1 3PJ, U.K.; email: justin.coon@eng.ox.ac.uk.}}
 
\maketitle

\begin{abstract}
Recent research has demonstrated the importance of boundary effects on the overall connection probability of wireless networks, but has largely focused on convex domains. We consider two generic scenarios of practical importance to wireless communications, in which one or more nodes are located outside the convex space where the remaining nodes reside. Consequently, conventional approaches with the underlying assumption that only line-of-sight (LOS) or direct connections between nodes are possible, fail to provide the correct analysis for the connectivity. We present an analytical framework that explicitly considers the effects of reflections from the system boundaries on the full connection probability. This study provides a different strategy to ray tracing tools for predicting the wireless propagation environment. A simple two-dimensional geometry is first considered, followed by a more practical three-dimensional system. We investigate the effects of different system parameters on the connectivity of the network though analysis corroborated by numerical simulations, and highlight the potential of our approach for more general non-convex geometries.
\end{abstract}

\section{Introduction}
Autonomous, self-organizing wireless relay networks have gained a considerable amount of research interest over recent years \cite{Haenggi2009}.  The potential of such networks to improve throughput and extend geographical coverage without necessarily requiring an increase in transmit power or bandwidth have resulted in them being attractive solutions for many modern wireless systems, including WiMAX and LTE-Advanced \cite{Fu2009}.  With similar objectives, such a network architecture has also been proposed for use in communications at $60$ GHz \cite{Qiao2010, Qiao2011} where severe path loss can significantly limit the transmission range of devices.  

One fundamental requirement of multi-hop relay or mesh networks, however, is connectivity, i.e., any node should either be directly connected to every other node or through a series of intermediate nodes.  Over recent years, a considerable amount of attention has been devoted to understanding the connectivity in such architectures (see, e.g., \cite{Gupta1998, Hong2003, Li2009, Mao2012} and references therein).  In the most simplistic case, the connection probability based on the unit disk model, in which nodes can communicate only if they are within a specified distance from each other, has been investigated.  In \cite{Gupta1998}, for instance, the critical power such that an ad hoc network is asymptotically connected with probability one is derived.  The transmitting range for connectivity in sparse networks is investigated in \cite{Santi2003}, while the connection probability in the presence of mobile nodes is presented in \cite{Santi2005}.

On the other hand, more practical models with channel fading have been studied in \cite{Bettstetter2005, Miorandi2008, Stuedi2005, Miorandi2005, Mao2012, Coon2012, Coon2012a}.  In \cite{Bettstetter2005, Stuedi2005}, the impact of log-normal shadowing on the connectivity is addressed.  Means of computing the node isolation probability and coverage for dense ad hoc networks in the presence of fast fading is presented in \cite{Miorandi2008}.  More recently, the impact of boundaries in dense networks have been investigated in \cite{Coon2012}, where it is shown that in the dense regime, the geometry of a system should not be neglected in analyzing the full connection probability.

Although the connectivity of random networks has been extensively studied, most existing work has considered only convex geometries in which the network resides.  In other words, there is always an assumption that a connection between two nodes is only possible if a line of sight (LOS) exists between the two.  In this paper, we study the scenarios where such assumptions need not hold.  More specifically, we investigate the connectivity in a network where more than one node may reside outside the main system and close to gaps in the boundaries of the confining geometry.  Moreover, considering that in practical situations, wireless connectivity in confined geometries may involve reflections of signals from boundaries, we present a means of modeling these reflected rays using the principles from mathematical billiards \cite{NikolaiChernov2006}.  While reflections from other obstacles may occur in practice, these are less dominant compared to reflections from large, smooth surfaces such as walls.  On that account, these minor reflections are not explicitly considered in the derivations presented herein.  Rather, they are accounted for by choosing an appropriate statistical channel fading model.  

Two scenarios are considered in this contribution.  In the first case, the full connection probability of a network where one node (a receiver) is located outside a gap (the so-called `keyhole phenomenon') while all other transmitter nodes are internal to the main system is considered.  Under the system assumptions, connection from the outside node to any other node may only be possible through reflections.  In particular, this would be the case if the size of the gap is small and no LOS is present between the external and internal nodes.  We refer to this problem as the \textit{escape problem}, analogous to the problem in mathematical billiards where a particle escapes a given geometry through a gap in the boundary \cite{Dettmann2009}.  In the second scenario, an extension of the escape problem is investigated, where both transmitter and receiver nodes are positioned next to gaps outside the main system.  This problem is referred to as the \textit{transport problem}, analogous to the work done in \cite{Dettmann2011}.  From a wireless communications perspective, such scenarios are particularly relevant for transmission at high frequencies where communication through walls is likely to render the signal undetectable at the receiver.  One potential application is the transmission at 60 GHz.  Throughout this work, we will assume that communication will be at high frequencies.  Nevertheless, it should be borne in mind that the analysis presented herein is applicable to any range of transmission frequency and conclusions drawn in this paper may be applicable to other systems as well.

For the reader's benefit and to motivate our work, we would like to draw a comparison between using the analytical tools from connectivity theory and the conventional approach to modeling a wireless communication environment.  Modeling and predicting such environments have been very actively researched over the last decades (see, e.g., \cite{Young1996, Yang1998, Mackay1999, Zhang2000, Zhang2002, Choudhury2011} and references therein).  Conventionally, ray tracing tools and software have been used for the characterization of the received signal strength at given points in different scenarios.  For instance, ray tracing techniques for underground mines and monorails have been addressed in \cite{Choudhury2011}.  Such tools, however, are generally based on the model where there is only one transmitting device whose location is known a priori.  By factoring in the boundaries of the system, reflection, diffraction, and absorption coefficients among other parameters, an estimate of the received signal strength can be obtained through a simulation-based technique.  In more complex scenarios where multiple nodes may be transmitting concurrently, such as mesh or ad hoc networks, using such an approach may be prohibitively complex.  The analysis of connectivity presented herein, on the other hand, is based on an analytical framework that allows the system designer to quantify the performance in any given environment, regardless of the node locations.  Generally, the analytics can be very easily visualized using common mathematical software such as Mathematica or Matlab.  This set of tools is more suitable when there is no fixed infrastructure.  It should be borne in mind that the authors are not arguing about the effectiveness of ray tracing software in the design of wireless systems.  Rather, we aim at encouraging the use of connectivity theory to obtain a higher level analysis of large scale network properties.  This analysis can be used alongside ray tracing software in optimizing a wireless system.

This paper is structured as follows.  In Section \ref{sec:escape}, we present the general setting for the escape problem and derive the full connection probability of the network.  We also present some analytical results for this scenario.  In Section \ref{sec:transport}, we extend our analysis to the transport problem.  Finally some concluding remarks are presented in Section \ref{sec:conclusion}.  To further aid the reader, we provide a glossary of the most relevant system parameters in Table \ref{tab:glossary}.
\begin{table}[t]
\caption{Glossary of Common Symbols}
\centering
\begin{tabular}{| l | p{8cm} |}
\hline\hline
\textbf{Symbol} & \textbf{Description} \\[0.5ex]
\hline
$\mathcal{A}$ & Set of nodes\\
$\alpha$ & Signal attenuation factor upon collision  \\
$\beta$  & Dimensionless constant depending on propagation environment \\
$\phi$   & Escape angle \\
$\theta$ & Maximum escape angle\\
$\eta$   & Path loss exponent \\
$\nu$ and $\mu$ & Parameters dependent on the Rice $K$-factor when approximating the Marcum $Q-$function to an exponential function, c.f., \cite{Bocus2013} \\
$c$ & Reflection index \\
$C$ & Maximum number of signal reflections considered \\
$\mathcal{D}_i$ & Region covered by $i$ signal reflections \\
$\epsilon$ & Length of gap on the boundary \\
$G^{\mathcal{A}}$ & Set of graphs with nodes in $\mathcal{A}$ \\
$G^{\mathcal{A}}_j$ & Set of graphs with nodes in $\mathcal{A}$ having largest component of size $j$\\
$H_{ij}$ & Probability that nodes $i$ and $j$ are directly connected based on an SNR threshold, c.f., (\ref{eq:Hij_def}) \\
$L$ & Length of system boundary for a rectangular domain \\
$K$ & Rice factor \\
$w$ & Width of system for a rectangular domain \\[1ex]
\hline
\end{tabular}
\label{tab:glossary}
\end{table}

\section{The Escape Problem}
\label{sec:escape}
In this section, we analyze the \textit{escape problem}.  We start by describing the general settings of the system and present the means of analyzing the effects of reflections arising from the boundaries using some of the tools from mathematical billiards.  The full connection probability is derived next and finally some analytical results are presented.

\subsection{System Geometry}
\label{sec:system_geometry}
\begin{figure}[t]
	\centering
		\includegraphics[width=8cm]{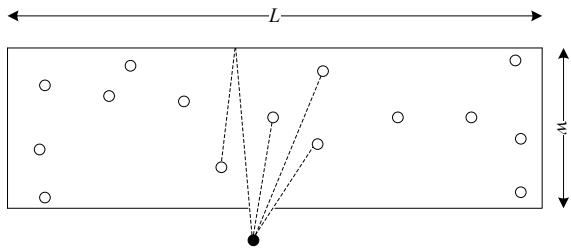}
	\caption{System model for the `escape problem', where one external node needs to connect to at least one internal node.}
	\label{fig:generic_model_escape}
\end{figure} 
Consider a simple system as shown in Fig. \ref{fig:generic_model_escape}, where a total of $N$ nodes are randomly distributed within the system boundaries and one node is located next to a gap outside the system.  Let the width and length of the rectangular bounding domain be $w$ and $L$ units respectively, where we assume that $L \gg w$ and the size of the gap, $\epsilon$, is much smaller than the other system dimensions.  For this system, we are interested in finding the probability that all the $N+1$ nodes are connected to each other.  This problem can be split into two parts, namely the probability that all nodes enclosed by the boundaries are fully connected and the probability that the external node is connected to at least one other node on the interior by a bridging link.  While some works have investigated the first part of the problem, the second part is still an unexplored issue.  Connection between the external node and any other node would be possible only if a signal path exists that can `escape' the main system to connect to the outside node.  This problem is similar to the work in \cite{Dettmann2009} where the authors investigate the probability that a particle can escape a stadium billiard through a gap along the boundary.  It should be noted that although a single external node is considered in this contribution for illustration purposes, the work can be readily extended to cater for multiple nodes, say $M$, that lie outside the system.  In such a case, the probability that a bridging link exists, i.e., one node within the system boundaries is connected to at least one external node would simply be given by one minus the probability that no bridging link exists raised to the power of $M$.  This can be expressed as $1-\left(1-\langle P_{b,k} \rangle \right)^M$, where $P_{b,k}$ is the probability that the external node $k$ is connected to one internal node and $\langle I \rangle$ refers to the average of the quantity $I$ over all configurations.  For the case of a single external node, the probability that a bridging link exists is simply given by $\langle P_{b,0} \rangle $, where the node $0$ is the external node.

\subsection{Modeling Reflections}
On the assumption of a very high node density in the enclosed space, an approach similar to that presented in \cite{Coon2012} can be followed in the derivation of the full connectivity probability of the network.  However, in situations where nodes are located close the gap inside the system, or alternately, an LOS transmission is not possible between the external and one of the internal nodes, it is more appropriate to study the problem using a different approach.  More specifically, the contribution of signals reflected from the boundaries should be considered in the connectivity analysis.  We aim to demonstrate the benefits of considering signal reflections under such conditions.  To this end, a proper description of the different reflected paths after a given number of reflections is required.  In particular, for a given external node location, the region that can be connected given a minimum number of reflections needs to be defined.

\begin{figure}[t]
	\centering
		\includegraphics[width=6cm]{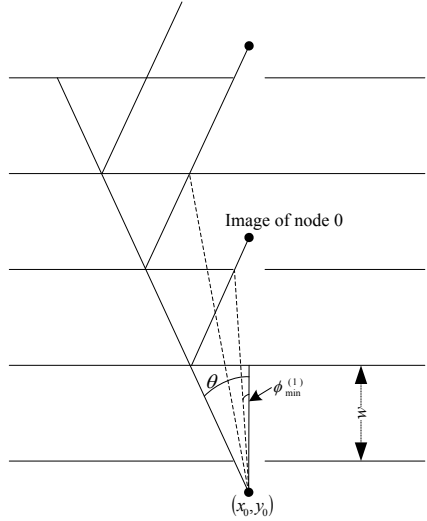}
	\caption{Illustration of how the method of unfolding can be used to determine the regions of interest for a given number of reflections.}
	\label{fig:reflection_region_nonoverlapping}
\end{figure}

Assume that the gap in the lower boundary of the system, as illustrated in Fig. \ref{fig:generic_model_escape}, is small and far from the vertical boundaries.  In that case, only reflections from the horizontal boundaries need to be considered.  Given the geometry and the non-chaotic behavior of the reflected signals, the method of trajectory unfolding can be employed to determine the reflected signal paths \cite{NikolaiChernov2006}.  The trick in this method consists in reflecting the system (rectangle in our case, c.f., Fig. \ref{fig:generic_model_escape}) instead of reflecting the trajectory of the signal.  Consequently, the signal path can be viewed as a single, straight line that travels through multiple images of the system enclosure.  An illustration of this concept is given in Fig. \ref{fig:reflection_region_nonoverlapping}, where the external node is referred to as node 0, with coordinates $(x_0,y_0)$.  We let the $y$-coordinate of the lower boundary be zero.  In this contribution, we assume that node 0 is fixed.  From a practical point of view, such an assumption is valid in scenarios where an external device is connected to some fixed or mobile infrastructure, such as a fixed base station installation on the exterior of an office block.

To determine the regions of interest, the geometry is analyzed using the polar coordinate system, in which case every region can be described by a range of escape angles from the gap and a range of distances.  The escape angle, $\phi$, is defined as the angle between the vertical axis and the signal trajectory.  We define the maximum escape angle as $\theta$, as shown in Fig. \ref{fig:reflection_region_nonoverlapping}.  It should be noted that although the representation indicates only the regions to the left of node 0, the process can be readily repeated to cover the regions on the right.  For the purpose of illustration, however, the region to the right of the external node will not be explicitly considered in what follows.  

Although any points within the boundaries can be connected by different paths undergoing different numbers of reflections, we are primarily interested in the strongest path in the analysis of the connectivity.  In general, considering the attenuation of a signal upon reflection with an object, the strongest path is more likely to be that which undergoes the least number of reflections.  On that account, we define the various regions within the boundaries based on the minimum number of reflections necessary for a signal to reach these.  For example, we define the region covered only by LOS paths (no reflections) as $\mathcal{D}_0$, while the region covered by one signal reflection, excluding the LOS region, is $\mathcal{D}_1$.  For a given number of reflections, $c$, $\mathcal{D}_c$ is determined by $\phi$ and $r$ which are defined as 
\begin{eqnarray}
\phi_{\min}^{(c)} \leq &\phi& \leq \theta \\
r_{\min}^{(c)}(\phi) \leq &r(\phi)& \leq r_{\max}^{(c)}(\phi)
\label{eq:rmin}
\end{eqnarray}
where 
\begin{eqnarray}
\phi_{\min}^{(c)} &=& \left\{\begin{array}{l l} 0, & c=0 \\
\tan^{-1} \frac{\left( (c-1)w +|y_0|\right)\tan\theta}{(c+1)w+|y_0|}, & c>0 \end{array}\right. \\
r_{\min}^{(c)}(\phi) &=& \left\{\begin{array}{l l} \frac{|y_0|}{\cos \phi}, & c =0 \\ \label{eq:rmin2}
\frac{2\left( cw+ |y_0| \right)\sin \theta}{\sin(\theta+\phi)}, & c>0 \end{array} \right. \\
r_{\max}^{(c)}(\phi) &=& \frac{(c+1)w + |y_0|}{\cos\phi}
\label{eq:rmax}
\end{eqnarray}
It should be noted that the value of $r$ is a function of the angle $\phi$
\begin{figure}[t]
	\centering
		\includegraphics[width=8cm]{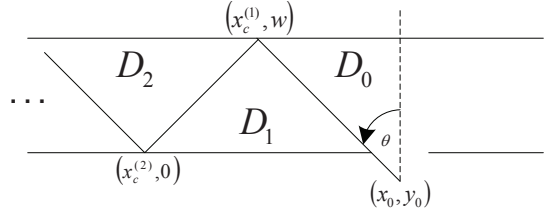}
	\caption{Illustration of the regions covered by different numbers of signal reflections with the boundaries.}
	\label{fig:reflection_cartesian}
\end{figure} 
For completeness, we also define the respective regions, $\mathcal{D}_c$, in the Cartesian coordinate system.  Let $x_c^{(i)}$ represents the point of impact of the signal on the horizontal boundaries after $i-1$ reflections for $1\leq i \leq c+1$, when the signal escapes the gap with the angle $\theta$ as shown in Fig. \ref{fig:reflection_cartesian}.  This value can be computed as
\begin{equation}
x_c^{(i)} = x_0 - \left(iw +|y_0| \right) \tan \theta
\end{equation}
It can be verified that the region to the left of node 0, $\mathcal{D}_0$, is bounded horizontally by $x = -(y-w)\tan\theta + x_c^{(1)}$ and $x = x_0$ and vertically by $y=0$ and $y=w$.  Similarly, region $\mathcal{D}_1$ is bounded by $x= (y-w)\tan\theta + x_c^{(1)}$ and $x = -(y-w)\tan\theta + x_c^{(1)}$, while $\mathcal{D}_2$ is bounded by $x = -(y-w)\tan\theta +x_c^{(3)}$ and $x= (y-w)\tan\theta + x_c^{(1)}$.  The definition of the boundaries of other regions follows the same approach.

\subsection{Probability of Full Connectivity}
For the derivation of the full connection probability, a cluster expansion technique similar to that presented in \cite{Coon2012} is followed.  

\subsubsection{First Order Cluster Expansion}
We define the set of nodes as $\mathcal{S}=\{0,1,\cdots,N \}$, where as previously stated node 0 is the external node and is fixed while the remaining nodes are randomly placed within the system.  We refer to the probability that two nodes $i$ and $j$, separated by a distance $\mathbf{d}(\mathbf{r}_i, \mathbf{r}_j)$, are connected as $H_{ij}$, where $\mathbf{r}_i$ is the coordinates of node $i$.  This pair connectedness probability is dependent on the signal-to-noise ratio (SNR), channel gains and other system parameters that will be introduced later.  In this work, we use the same notations as in \cite{Coon2012}, where a graph $g = (\mathcal{A},L)$ consists of a set $\mathcal{A} \subseteq \mathcal{S}$ and a collection of direct links $L\subseteq \{(i,j) \in\mathcal{A}:i,j \}$.  The set of graphs with nodes in $\mathcal{A}$ is defined by $G^{\mathcal{A}}$, while the set of graphs with nodes in $\mathcal{A}$ having largest connected component of size $j$ is expressed as $G^{\mathcal{A}}_j$.

For any pair of nodes, the trivial probability that they are connected or not is given by
\begin{equation}
1 \equiv H_{ij} +\left(1-H_{ij}\right).
\label{eq:Hij_def}
\end{equation}
Multiplying over all links in a set $\mathcal{A}$ to determine the probability of all possible combinations leads to 
\begin{equation}
1 = \prod_{i,j\in A;i<j} \left[H_{ij}+\left(1-H_{ij} \right) \right] = \sum_{g\in G^{\mathcal{A}}}\mathcal{H}_g
\end{equation}
where
\begin{equation}
\mathcal{H}_g = \prod_{(i,j)\in g} H_{ij} \prod_{(i,j)\notin g}  \left(1-H_{ij}\right).
\end{equation}
Setting $\mathcal{A} = \mathcal{S}$ yields
\begin{eqnarray}
1 &=& \underbrace{\sum_{g\in G_{N+1}^{\mathcal{S}}} \mathcal{H}_g}_{P_{fc}} + \sum_{g \in G_N{^\mathcal{S}}} \mathcal{H}_g  + \cdots + \sum_{g\in G_1{^\mathcal{S}}}\mathcal{H}_g
\end{eqnarray}
By taking the first order approximation of the above expression, we obtain
\begin{align}
P_{fc}& \approx 1 - \left\langle \sum_{g\in G_N{^\mathcal{S}}} \mathcal{H}_g \right\rangle \nonumber\\
&= 1- \left\langle\sum_{p=0}^N \prod_{j\neq p}\left(1-H_{jp} \right) \left(\sum_{g\in G_N^{\mathcal{S}\backslash\{p\}}} \mathcal{H}_g\right) \right\rangle \nonumber\\
&= 1 - \left\langle \prod_{j=1}^N \left(1-H_{0j} \right) \underbrace{\sum_{g\in G_N^{\mathcal{S}\backslash\{0\}}} \mathcal{H}_g}_{\approx 1} \right\rangle - \sum_{p=1}^N \left\langle\left(1- H_{0p} \right)\prod_{j=1,\cdots,N,j\neq p}\left(1-H_{jp}\right) \sum_{g\in G_{ N_0  }^{\mathcal{S}\backslash\{p \}}} \mathcal{H}_g \right\rangle. 
\label{eq:full_pc}
\end{align}
where $N_0 = N$.  The reason for the different notation here is to emphasize that the largest component of size $N$ in the last summation includes the external node 0.  Note that for $N$ randomly distributed nodes with locations $\mathbf{r}_i \in \mathcal{V}\subseteq \mathbb{R}^D$, $i=1,\cdots, N$, the average of a quantity $I$ over all configurations, $\langle I \rangle$,  is defined as
\begin{equation}
\langle I \rangle = \frac{1}{V^{N}} \int_{\mathcal{V}_N} I\left(\mathbf{r}_1, \mathbf{r}_2, \cdots, \mathbf{r}_N \right) d\mathbf{r}_1\cdots d\mathbf{r}_N
\end{equation}
where $V = |\mathcal{V}|$.

The last expression in (\ref{eq:full_pc}) indicates that $\left\langle\left(\sum_{p=0}^N \prod_{j\neq p}\left(1-H_{jp} \right) \right)\left(\sum_{g\in G_N^{S\backslash\{p\}}} \mathcal{H}_g\right) \right\rangle$ is made up to two terms; the first corresponds to the case where all nodes within the system boundaries are connected to each other but not to node 0, while the second term represents the case where node $j, j\geq 1$, is isolated while node $0$ is connected to at least one node on the interior.  Each of these terms is analyzed next.

Consider the first average term in (\ref{eq:full_pc}).  We refer to this term as the \textit{exterior node isolation probability}.
\begin{align}
&\left\langle  \prod_{j=1}^N \left(1-H_{0j} \right) \underbrace{\sum_{g\in G_N^{S\backslash\{0\}}} \mathcal{H}_g}_{\approx 1}  \right\rangle \approx \left \langle\prod_{j=1}^N \left(1-H_{0j} \right) \right\rangle \nonumber\\
= & \frac{1}{V^N} \int_{\mathcal{V}^N} \prod_{j=1}^N \left( 1- H_{0j}\right) d\mathbf{r}_1 \cdots d\mathbf{r}_N \nonumber\\
= & \left[\frac{1}{V} \int_{\mathcal{V}} \left(1-H_{01} \right)d\mathbf{r}_1 \right]^N\nonumber\\
= & \left[1-\frac{1}{V}\int_{\mathcal{V}} H_{01}d\mathbf{r}_1 \right]^N \approx e^{-\rho \int_{\mathcal{V}} H_{01}d\mathbf{r}_1}\left( 1 + \mathcal{O}\left(\frac{\rho^2}{N} \right) \right)
\label{eq:H01}
\end{align}
where $\rho =\frac{N}{V}$ represents the density of the system and the last approximation holds in a homogeneous network and for large $N$.  The interested reader is referred to equations (7) and (8) in \cite{Coon2012} for further discussion on this approximation.  Under the same assumptions, the second summation in (\ref{eq:full_pc}) becomes 
\begin{align}
\sum_{p=1}^N & \left\langle\left(1- H_{0p} \right)\prod_{j=1,\cdots,N,j\neq p}\left(1-H_{jp}\right) \sum_{g\in G_{N_0}^{S\backslash\{\rho \}}} \mathcal{H}_g \right\rangle \nonumber\\
&\approx N\left\langle \left(1-H_{0N} \right)\prod_{j=1}^{N-1} \left(1- H_{jN} \right) \right\rangle \nonumber\\
&= \frac{N}{V^{N}} \int_{\mathcal{V}^{N}} \left(1-H_{0N}\right)\prod_{j=1}^{N-1} \left(1-H_{jN} \right) d\mathbf{r}_1\cdots d\mathbf{r}_N \nonumber\\
&= \frac{N}{V}\int_{\mathcal{V}} \left(1-H_{0N}\right)\left(1-\frac{1}{V}\int_{\mathcal{V}} H_{1N} d\mathbf{r}_1 \right)^{N-1} d\mathbf{r}_N\nonumber\\
&\approx \rho \int_{\mathcal{V}} e^{-\rho\int_{\mathcal{V}} H_{1N}d\mathbf{r}_1} d\mathbf{r}_N - \rho \int_{\mathcal{V}} H_{0N}\left(1-\frac{1}{V}\int_{\mathcal{V}} H_{1N}d\mathbf{r}_1 \right)^{N-1} d\mathbf{r}_N
\label{eq:H_1N_express}
\end{align}
where the approximations in (\ref{eq:H_1N_express}) follow the same procedures as in (\ref{eq:H01}) \cite{Coon2012}.  Depending on the system configuration, either of the average terms in (\ref{eq:full_pc}) can have a more pronounced effect on the full connection probability.  Under the assumptions of a high internal node density and a small gap in the boundary, the probability that all nodes being connected would be mostly influenced by whether the external node is connected to at least one other node.  Consequently, the contribution of (\ref{eq:H_1N_express}) on the full connectivity of the network would be negligible.  We analyze such scenarios in this contribution and focus on the term in (\ref{eq:H01}).

\subsubsection{Pair-Connected Functions and Connectivity Mass}
In this subsection, we will use the recently established notion of connectivity mass of the pair connectedness function, $H_{01}$, for the analysis of the full connection probability \cite{Coon2012}.  Typically, the connectivity mass is defined as the integral of the pair connectedness function over all configurations.  The pair connectedness function in the scenario under investigation can be viewed as the complement of the information outage probability, $P_{out}$, which can be expressed as \cite{Proakis2000}
\begin{eqnarray}
P_{out} &=& P\left(\log_2(1+\text{SNR}\times |h|^2) <R_0 \right) \nonumber\\
&=& P\left( |h|^2 <\frac{2^{R_0}-1}{\text{SNR}}  \right)
\end{eqnarray}
where $|h|^2$ is the channel gain between two nodes, $R_0$ is the minimum information rate and the signal-to-noise ratio is a quantity that is proportional to
\begin{equation}
\text{SNR} \propto G_T G_R r^{-\eta} \alpha^{c}
\end{equation}
where $G_T$ and $G_R$ are the antenna gains at the transmit and receive nodes respectively which is unity under isotropic radiation, $r$ is the dimensionless distance (relative to the signal wavelength) between the connected nodes, $\alpha$ is the dimensionless factor representing the amount of signal attenuation upon collision/reflection with the boundaries \cite{Zhao2001}, $c$ is the number of reflections in the signal trajectory and $\eta$ is the environment dependent path loss exponent.  Typical values of $\eta$ are $2$ under the free space path loss assumption and $\eta>2$ for dense or crowded environments.  In the previous work analyzing wireless connectivity in confined geometries \cite{Coon2012}, a fair assumption was made that a Rayleigh fading channel is present and the random variable $|h|^2$ is exponentially distributed.  However, since we are considering reflected signals as well in this paper, a more appropriate model for the channel fading would be the Rician fading model \cite{Proakis2000}.  Such a fading model is particularly suitable when a strong signal can be observed at the receiver along with other weaker signals.  The stronger signal typically is the one that travels the shortest path, and equivalently, undergoes fewer reflections before reaching the destination node.

The channel power density of a Rician fading channel is given by \cite{Proakis2000}
\begin{equation}
f_X(x) = \omega^{-1} (K+1) e^{-(K+\omega^{-1}(K+1)x)}I_0\left(\sqrt{\frac{4K(K+1)x}{\omega}} \right)
\end{equation}
where $K$ is the Rice factor, generally taken to be greater than unity (with typical values around 4 or 5 being reasonable for this work), $\omega$ is a channel dependent parameter and $I_0$ is the modified Bessel function of the first kind.  The cumulative distribution function is 
\begin{equation}
F_X(x) = 1 - Q_1\left(\sqrt{2K},\sqrt{\frac{2(K+1)x}{\omega}} \right)
\end{equation}
where $Q_1(a,b)$ is the Marcum $Q$-function.  It follows that the connection probability for a pair of nodes is 
\begin{eqnarray}
H &=& 1-P_{out} \nonumber\\
&=& Q_1\left(\sqrt{2K},\sqrt{2(K+1) \beta r^{\eta} \alpha^{-c}} \right)
\end{eqnarray}
where $\beta$ is a proportionality constant dependent on parameters such as frequency of transmission, transmit power and system scaling.  For practical situations, it is often the case that $\beta \ll 1$.  

Let us first consider the connectivity mass function involving the $H_{01}$ term.  With reference to (\ref{eq:H01}) and assuming a path loss exponent of $\eta = 2$ (free space path loss),
\begin{equation}
\int_{\mathcal{V}} H_{01} d\mathbf{r} = \sum_{c=0}^C \int_{\phi_{\min}^{(c)}}^\theta \int_{r_{\min}^{(c)}(\phi)}^{r_{\max}^{(c)}(\phi)} r Q_1\left(\sqrt{2K},\sqrt{2(K+1)\beta r^2 \alpha^{-c}} \right) dr d\phi 
\end{equation}
It should be noted that the above expression is valid for the two-dimensional rectangular domain presented earlier.  Similar techniques can be followed for other geometries. 

\begin{figure}[t]
	\centering
		\includegraphics[width=9cm]{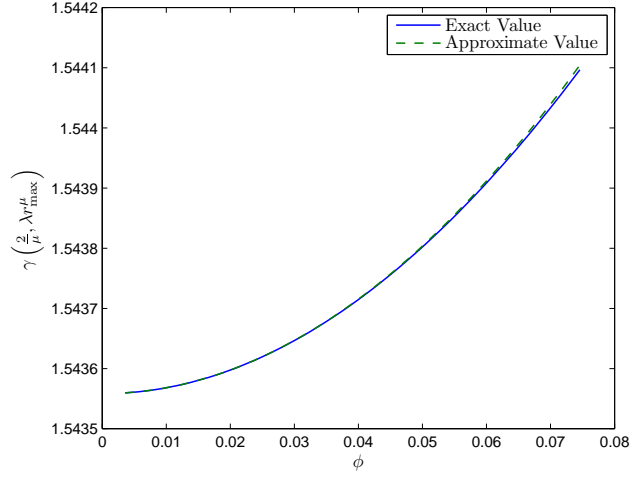}
	\caption{Comparing $\gamma\left(\frac{2}{\mu},\lambda_c r_{\max}^{\mu} \right)$ with the approximation in (\ref{eq:approx_lower_inc_rmax}) for $c=1$, $K=4$, $\beta = 10^{-3}$, $w=20$, $L=100$, $\epsilon = 0.3$, $y_0 = -2$, $\theta=\tan^{-1}\left(\frac{\epsilon}{2y_0}=0.0749 \right)$.}
	\label{fig:compare_incgamma_rmax} 
\end{figure}

\begin{figure}[t]
	\centering
		\includegraphics[width=9cm]{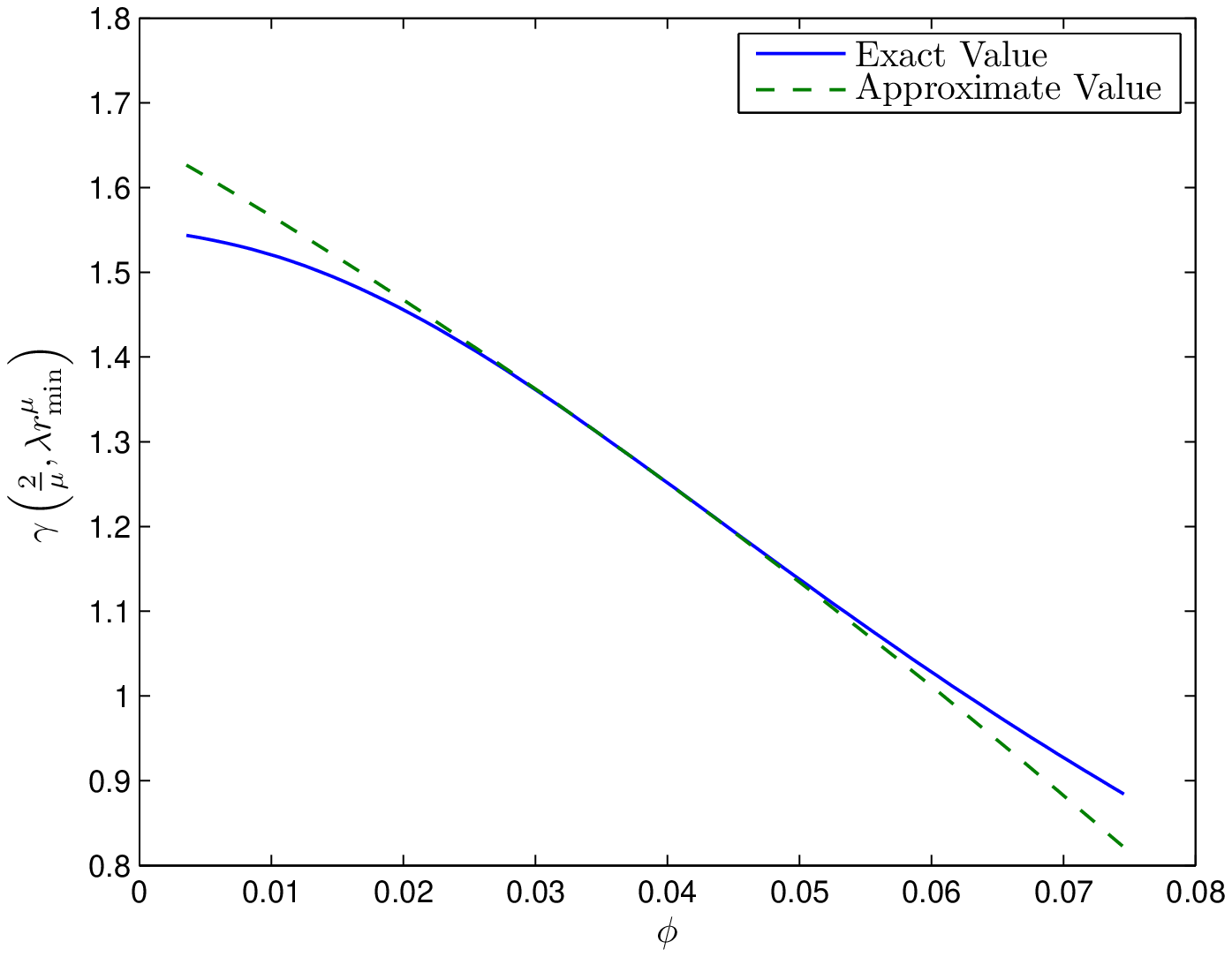}
	\caption{Comparing $\gamma\left(\frac{2}{\mu},\lambda_c r_{\min}^{\mu} \right)$with the approximation in (\ref{eq:approx_lower_inc}) for $c=1$, $K=4$, $\beta = 10^{-3}$,$w=20$, $L=100$, $\epsilon = 0.3$, $y_0 = -2$, $\theta=\tan^{-1}\left(\frac{\epsilon}{2y_0}=0.0749 \right)$.}
	\label{fig:compare_incgamma_rmin}
\end{figure}

To move forward in the analysis and perform the integration, we use the approximation of the Marcum $Q-$function presented in \cite{Bocus2013}, i.e., $Q_1(a,b)$ is approximated by $\exp\left(-e^{\nu} b^{\mu} \right)$ where $\nu$ and $\mu$ are parameters dependent on the parameter $a$, which in our case is equal to $\sqrt{2K}$.  It follows that 
\begin{align}
\int_{\mathcal{V}}& H_{01} d\mathbf{r} =  \sum_{c=0}^C \int_{\phi_{\min}^{(c)}}^{\theta} \int_{r_{\min}^{(c)}(\phi)}^{r_{\max}^{(c)}(\phi)} r e^{-\lambda_c r^{\mu}} dr d\phi \nonumber\\
= & \sum_{c=0}^C \int_{\phi_{\min}^{(c)}}^{\theta} \frac{\lambda_c^{-\frac{2}{\mu}}}{\mu} \left[\gamma\left(\frac{2}{\mu},\lambda_c (r_{\max}^{(c)}(\phi))^{\mu} \right) - \gamma\left(\frac{2}{\mu},\lambda_c (r_{\min}^{(c)} (\phi) )^{\mu} \right) \right] d\phi 
\label{eq:H_01_full}
\end{align}
where $\lambda_c = e^{-\nu} \sqrt{2(K+1)\beta\alpha^{-c}}$, $\gamma(.,.)$ is the lower incomplete gamma function and $C$ is the maximum number of reflections considered.  Typically, more than five collisions/reflections with obstacles would result in the signal being too heavily attenuated to be recovered, especially for high frequency transmission, such as at $60$ GHz.  It should be noted that in (\ref{eq:H_01_full}), only the region to the left of node 0 in Fig. \ref{fig:reflection_region_nonoverlapping} is considered for conciseness.  Extending the results for the general case is straightforward.  To enable the integration of the lower incomplete gamma function with respect to the parameter $\phi$, we consider its series expansion.  Referring to (\ref{eq:rmax}) and under the practical assumption that $\theta$ is small (a `keyhole' gap), the expansion about $\phi=0$ can be taken for the term involving $r_{\max}^{(c)}(\phi)$.  However, considering the definition of $r_{\min}(\phi)$ in (\ref{eq:rmin2}), it is realized that an expansion about $\phi=0$ may not be appropriate given that the cosecant function is not defined at the origin.  Instead, an expansion about $\frac{\theta}{2}$ is taken.  The corresponding expansion of the lower incomplete gamma functions are as expressed in (\ref{eq:approx_lower_inc_rmax}) to (\ref{eq:B_H01}) given below.
\begin{eqnarray}
\label{eq:approx_lower_inc_rmax}
\gamma\left(\frac{2}{\mu},\lambda_c \left((c+1)w+|y_0| \right)^{\mu} \sec^{\mu} \phi \right) &\approx& \gamma\left(\frac{2}{\mu},\lambda_c\left( (c+1)w+|y_0|\right)^{\mu} \right) +\nonumber\\
&& \frac{\mu}{2}\lambda_c^{\frac{2}{\mu}} \left( (c+1)w+|y_0|\right)^2e^{-\lambda_c((c+1)w+|y_0|)^{\mu}} \phi^2 +\mathcal{O}(\phi^4) \\
\gamma\left(\frac{2}{\mu}, \lambda_c\left(2(cw+|y_0|)\frac{\sin(\theta)}{\sin(\theta+\phi)} \right)^{\mu} \right) &\approx& \gamma\left(\frac{2}{\mu},2^{\mu} \lambda_c\left( (cw+|y_0|) \frac{\sin(\theta)}{\sin(3\theta/2)} \right)^{\mu} \right) - A \cot\left(\frac{3\theta}{2}\right)\left(\phi-\frac{\theta}{2} \right) -\nonumber\\
&& AB\left(\phi-\frac{\theta}{2} \right)^2 +\mathcal{O}\left(\phi-\frac{\theta}{2} \right)^3
\label{eq:approx_lower_inc}
\end{eqnarray}
where
\begin{eqnarray}
A &=& 4\mu \frac{\sin^2(\theta)}{\sin^2\left(\frac{3\theta}{2}\right)} (cw+|y_0|)^2 \lambda_c^{\frac{2}{\mu}} \exp\left(-2^{\mu}\lambda_c(cw+|y_0|)^{\mu}\frac{\sin^{\mu}(\theta)}{\sin^{\mu}(3\theta/2)} \right) \\
B &=& \frac{1}{4 \sin^2(3\theta/2)} \left(4^{\mu} \mu \lambda_c\left(\frac{(cw+|y_0|)\cos(\theta/2)}{1+2\cos(\theta)} \right)^{\mu} \left(1+\cos(3\theta) \right) - 2(2+\cos(3\theta)) \right).
\label{eq:B_H01}
\end{eqnarray}
To show the suitability of the expansions for the above integration, we plot the actual and approximated values of the corresponding lower incomplete gamma functions in Fig. \ref{fig:compare_incgamma_rmax} and \ref{fig:compare_incgamma_rmin}.  It can be observed that over the range of integration (c.f., (\ref{eq:H_01_full})), the approximated values are reasonably close to the actual ones.  Similar observations can be made for larger values of $\theta$ than those considered in the plot.

Substituting (\ref{eq:approx_lower_inc_rmax}) and (\ref{eq:approx_lower_inc}) into (\ref{eq:H_01_full}) leads to 
\begin{eqnarray}
\int_{\mathcal{V}} H_{01} d\mathbf{r} &\approx& \frac{\lambda_0^{-\frac{2}{\mu}}}{\mu} \int_0^{\theta} \left[\begin{array}{l}\gamma\left(\frac{2}{\mu},\lambda_0(w+|y_0|)^{\mu} \right) - \gamma\left(\frac{2}{\mu},\lambda_0|y_0|^{\mu} \right) + \\
\frac{\mu\phi^2}{2}\lambda_0^{\frac{2}{\mu}} \left((w+|y_0|)^2 e^{-\lambda_0(w+|y_0|)^{\mu}}-|y_0|^2e^{-\lambda_0|y_0|^{\mu}} \right)\end{array}\right] d\phi \nonumber\\
 +&& \sum_{c=1}^C \frac{\lambda_c^{-\frac{2}{\mu}}}{\mu} \int_{\phi_{\min^{(c)}}}^{\theta} \left[\begin{array}{l} 
\gamma\left(\frac{2}{\mu},\lambda_c\left((c+1)w+|y_0| \right)^{\mu} \right) +\frac{\mu \phi^2}{2}\lambda_c^{\frac{2}{\mu}}\left((c+1)w+|y_0| \right)^2 e^{-\lambda_c ((c+1)w+|y_0|)^{\mu}} -\\
\gamma\left(\frac{2}{\mu},2^{\mu}(cw+|y_0|)^{\mu}\lambda_c \frac{\sin^{\mu}(\theta)}{\sin^{\mu}(3\theta/2)} \right)+ \left(\phi-\frac{\theta}{2} \right)\cot(\frac{3\theta}{2})A +  AB\left(\phi-\frac{\theta}{2} \right)^2 \end{array}\right] d\phi \nonumber\\
&=& \frac{\lambda_0^{-\frac{2}{\mu}}}{\mu}\left[\begin{array}{l} \theta \left(\gamma\left(\frac{2}{\mu},\lambda_0(w+|y_0|^{\mu}) \right) - \gamma\left(\frac{2}{\mu},\lambda_0 |y_0|^{\mu} \right) \right) + \\  
\frac{\mu\theta^3}{6} \lambda_0^{\frac{2}{\mu}} \left( (w+|y_0|)^2 e^{-\lambda_0(w+|y_0|)^{\mu}} - |y_0|^2 e^{-\lambda_0 |y_0|^{\mu}} \right)
\end{array}\right] \nonumber\\
&+& \sum_{c=1}^C \frac{\lambda_c^{-\frac{2}{\mu}}}{\mu} \left[\begin{array}{l} \left(\theta - \phi_{\min}^{(c)}\right) \left( \gamma\left(\frac{2}{\mu},\lambda_c\left((c+1)w+|y_0| \right)^{\mu} - \gamma\left(\frac{2}{\mu},2^{\mu}(cw+|y_0|)^{\mu}\lambda_c \frac{\sin^{\mu}(\theta)}{\sin^{\mu}(3\theta/2)} \right)\right) \right) \\
+ \frac{A}{2}\cot\left(\frac{3\theta}{2}\right)\left(\theta^2 -\left(\phi_{\min}^{(c)}\right)^2 \right)+\frac{AB}{3} \left(\frac{\theta^3}{8} -\left(\phi_{\min}^{(c)} - \frac{\theta}{2} \right)^3 \right)\\
 + \frac{\mu}{6}\left( \theta^3 -(\phi_{\min}^{(c)})^3 \right)\lambda_c^{\frac{2}{\mu}}\left((c+1)w+|y_0| \right)^2 e^{-\lambda_c ((c+1)w+|y_0|)^{\mu}}
\end{array}\right].
\label{eq:mass_H01}
\end{eqnarray}
Recall that the first average term in (\ref{eq:full_pc}) corresponds to the exterior node isolation probability.  For the sake of completeness, we have presented the method of evaluating the second average term in (\ref{eq:full_pc}) in \ref{sec:appA}.

\subsection{Extension to the $3-$D Case}
\begin{figure}[t]
	\centering
		\includegraphics[width=9cm]{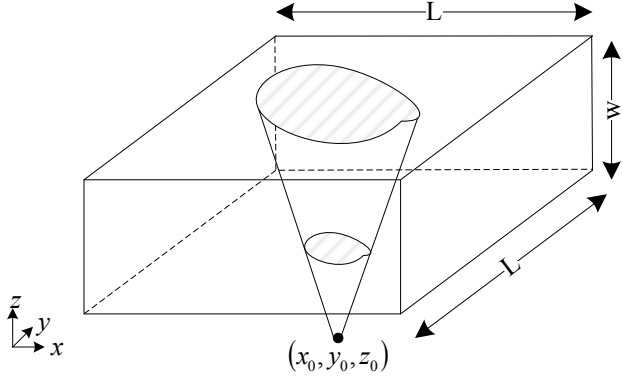}
	\caption{System model for 3-D case; the gap in the lower boundary can take any shape.}
	\label{fig:system_model_3D}
\end{figure}

In the previous subsection, a simple 2-D system was considered.  Nevertheless, extension of the analysis to more practical 3-D systems is straightforward, as will be shown here.  Consider a system model as shown in Fig. \ref{fig:system_model_3D} and let the gap in the lower boundary take any shape.  Similar to the 2-D case, we assume that the length and depth of the system, $L$, is much greater than the height $w$ for the same reasons as previously mentioned.  

In this subsection, we focus primarily on the probability of the external node being connected at least one internal node (c.f., the first average term in (\ref{eq:full_pc})).  To define the regions within the system boundaries covered by a given number of reflections from the boundaries, we make use of the spherical coordinate system such that any point within the geometry can be defined by three variables, namely, $r_{\phi}$, $\phi$ and $\varphi$.  The first variable, $r_{\phi}$, represents the distance from the external node to an arbitrary point inside the geometry; $\phi$, in this case, corresponds to the inclination with respect to the $z-$axis while the azimuth angle with respect to the $x-$axis is defined by the angle $\varphi$.  Considering a fixed external node at $(x_0,y_0,z_0)$, any point (in Euclidean space) inside the boundaries can be expressed as a function of this 3-tuple as
\begin{eqnarray}
x(r,\phi,\varphi) &=& x_0 +r_{\phi} \cos\left(\varphi\right) \sin\left(|\phi| \right) \nonumber\\ 
y(r,\phi,\varphi) &=& y_0 +r_{\phi} \sin\left(\varphi\right) \sin\left(|\phi| \right) \nonumber\\ 
z(r,\phi,\varphi) &=& z_0 +r_{\phi} \cos\left(|\phi| \right). 
\end{eqnarray}
Analogous to the 2-D case, the region $\mathcal{D}_c$ for $c$ reflections is defined by the following range of distances:
\begin{eqnarray}
r_{\min}^{(c)} \left(\phi,\varphi \right) &=& \left\{ \begin{array}{l l} \frac{|z_0|}{\cos \phi}, & c=0 \\
\frac{2\left(cw+|z_0|\right)\sin \left(\theta(\varphi) \right) }{\sin\left(\theta(\varphi) +\phi \right)}, & c >0 \end{array} \right. \nonumber\\
r_{\max}^{(c)}\left(\phi,\varphi \right) &=& \frac{(c+1)w + |z_0|}{\cos(\phi)}
\end{eqnarray}
while 
\begin{equation}
\phi_{\min}^{(c)}\left(\varphi \right) =\left\{ \begin{array}{l l} 0, & c =0 \\
\tan^{-1} \frac{\left( (c-1)w +|z_0| \right)\tan \left(\theta(\varphi) \right)}{(c+1)w + |z_0|}, & c>0\end{array}\right.
\end{equation}
Assuming a Rician fading channel as in the 2-D scenario, the connectivity mass (c.f., (\ref{eq:H_01_full})) can be evaluated by the integral
\begin{eqnarray} 
\label{eq:intH01_3D}
\int_{\mathcal{V}} H_{01} d \mathbf{r} &\approx& \sum_{c=0}^{C} \int_{0}^{2\pi} \int_{\phi_{\min}^{(c)}}^{\theta(\varphi)} \int_{r_{\min}^{(c)}}^{r_{\max}^{(c)}} \exp(- \lambda_c r^{\mu} ) r^{2} \sin{\phi} \ dr \ d\phi \ d\varphi \nonumber\\
&=& \sum_{c=0}^{C} \int_{0}^{2\pi} \int_{\phi_{\min}^{(c)}}^{\theta (\varphi)} \frac{\lambda_c^{-\frac{3}{\mu}}}{\mu} \left[ \gamma \left(\frac{3}{\mu}, \lambda_c \left(r_{\max} \right)^{\mu} \right) - \gamma \left(\frac{3}{\mu}, \lambda_c \left(r_{\text{min}}\right)^{\mu} \right) \right] \sin{\phi} \ d\phi \ d\varphi 
\end{eqnarray}
where $\lambda_c = e^{\nu} \kappa_c^{\mu}$.  Using similar approximations as in (\ref{eq:approx_lower_inc}) and assuming a circular gap on the lower face of the system, the integral evaluates to (\ref{eq:H_01_3D2}).
\begin{align}
\int_{\mathcal{V}} H_{01} & d\mathbf{r} \approx 2\pi \frac{\lambda_0^{-\frac{3}{\mu}}}{\mu}\left[\begin{array}{l} -\cos\left(\theta(\varphi) \right) \left(\gamma\left(\frac{3}{\mu},\lambda_0(w+|z_0|^{\mu}) \right) - \gamma\left(\frac{3}{\mu},\lambda_0 |z_0| \right)^{\mu} \right) + \\  
\left(-2 - (-2+ \theta(\varphi)^2)\cos(\theta(\varphi)) + 2\theta(\varphi) \sin(\theta(\varphi)) \right) \frac{\mu}{2} \lambda_0^{\frac{3}{\mu}} \times \\
\left( (w+|z_0|)^2 e^{-\lambda_0(w+|z_0|)^{\mu}} - |z_0|^2 e^{-\lambda_0 |z_0|^{\mu}} \right)
\end{array}\right] \nonumber\\
&+ 2\pi \sum_{c=1}^C \frac{\lambda_c^{-\frac{3}{\mu}}}{\mu} \left[ \begin{array}{l}
\left(\gamma\left(\frac{3}{\mu}, \lambda_c \left((c+1)w+|z_0|\right)^{\mu} \right) -\gamma\left(\frac{3}{\mu}, 2^{\mu} \lambda_c (cw+|z_0|)^{\mu} \frac{\sin^{\mu} (\theta(\varphi))}{\sin^{\mu}\left(\frac{3\theta(\varphi)}{2} \right)} \right) \right)\\
\left(\cos(\phi_{\min}^{(c)}) - \cos(\theta(\varphi) ) \right) \\
+\frac{\mu}{2}\lambda_c^{\frac{3}{\mu}}\left((c+1)w +|z_0| \right)^2 e^{-\lambda_c\left((c+1)w+|z_0| \right)^{\mu}} C_{3d} +\cot(3\theta(\varphi)/2)A_{3d}D_{3d} + A_{3d} B_{3d}E_{3d} 
\end{array}\right]\nonumber\\
\label{eq:H_01_3D2}
\end{align} 
where
\begin{eqnarray}
A_{3d} &=& 8\mu \frac{\sin^3(\theta(\varphi))}{\sin^3\left(\frac{3\theta(\varphi)}{2}\right)} (cw+|z_0|)^3 \lambda_c^{\frac{3}{\mu}} \exp\left(-2^{\mu}\lambda_c(cw+|z_0|)^{\mu}\frac{\sin^{\mu}(\theta(\varphi))}{\sin^{\mu}(3\theta(\varphi)/2)} \right) \nonumber\\
B_{3d} &=& \frac{1}{2}\left(\cot^2\left(\frac{3\theta(\varphi)}{2} \right) \left( 4^{\mu} \mu \lambda_c\left(\frac{(cw+|z_0|)\cos(\theta(\varphi)/2)}{1+2\cos(\theta(\varphi))} \right)^{\mu} -4 \right) -1 \right) \nonumber\\
C_{3d} &=& -\left(\theta(\varphi)^2 -2 \right)\cos\left(\theta(\varphi) \right) + \left((\phi_{\min}^{(c)})^2-2 \right)\cos\left(\phi_{\min}^{(c)} \right) + 2\theta(\varphi) \sin\left( \theta(\varphi) \right) - 2\phi_{\min}^{(c)} \sin\left(\phi_{\min}^{(c)} \right) \nonumber\\
D_{3d} &=& -\frac{\theta(\varphi)^2}{2} \cos\left( \theta(\varphi) \right) +\left(\phi_{\min}^{(c)} - \frac{\theta(\varphi)}{2} \right)\cos\left(\phi_{\min}^{(c)} \right) + \sin\left( \theta(\varphi) \right) - \sin\left( \phi_{\min}^{(c)}\right) \nonumber\\
E_{3d} &=& -\frac{1}{4}\left(\theta(\varphi)^2-8 \right) \cos\left( \theta(\varphi) \right) +\frac{1}{4} \left( -8 +(\theta(\varphi) - 2\phi_{\min}^{(c)})\right)\cos\left( \phi_{\min}^{(c)}\right) + \theta(\varphi) \sin\left(\theta(\varphi)\right) \nonumber\\
&& +\left( \theta(\varphi) - 2\phi_{\min}^{(c)}\right)\sin(\phi_{\min}^{(c)}).
\end{eqnarray}

\subsection{Numerical Results and Discussion}
\subsubsection{2-D Scenario}
\begin{figure}[t]
	\centering
		\includegraphics[width=9cm]{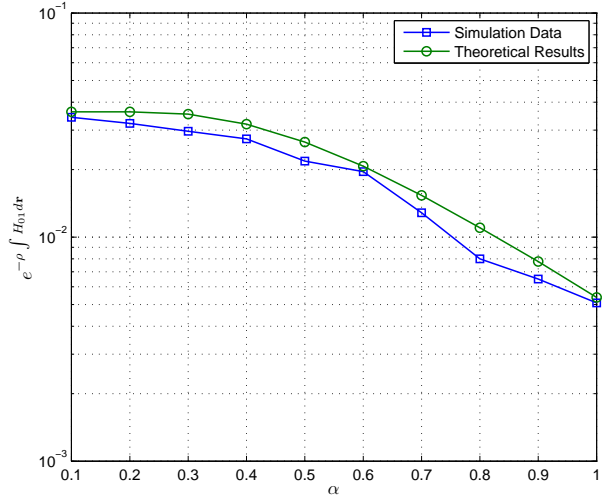}
	\caption{Comparison of theoretical and simulation results of the external node isolation probability, $\epsilon = 0.3$, $w=20$, $L=100$, $y_0=-2$, $\rho=0.1$, $\theta = \tan^{-1}\left(\frac{\epsilon}{2y_0}\right)=0.0749$.}
	\label{fig:compare_simul_theory_escapa_out}
\end{figure}

We first compare the theoretical results in (\ref{eq:mass_H01}) with simulation results for this function plotted against $\alpha$ in Fig. \ref{fig:compare_simul_theory_escapa_out}.  For the purpose of this simulation, we consider a simple example where the dimension of the boundaries are $w=20$ and $L=100$ units and the size of the gap is $\epsilon = 0.3$ units.  As described in Section \ref{sec:system_geometry}, we assume that the horizontal positioning of node $0$ is half-way between the gap and we set $y_0=-2$.  With these parameter values, $\theta$ is seen to be equal to $\tan^{-1} \left(\frac{0.15}{2}\right)$ radians.  Furthermore, $\rho=0.1$ and the system dependent parameter is chosen to be $\beta=10^{-3}$ in this example.  It should be noted that although the value chosen for $\beta$ is arbitrary, this value is of the same order as that which would be obtained with practical values of transmit power, transmission frequency and average fade margin.  The simulated results were obtained by finding the number of occurrences over which node $0$ could not connect to any other nodes while all nodes inside the system were fully connected over $10000$ different realizations of the channels and node locations.  The closeness of the theoretical and simulated results validates the derivation of the connectivity mass above.

\begin{figure}[t]
	\centering
		\includegraphics[width=9cm]{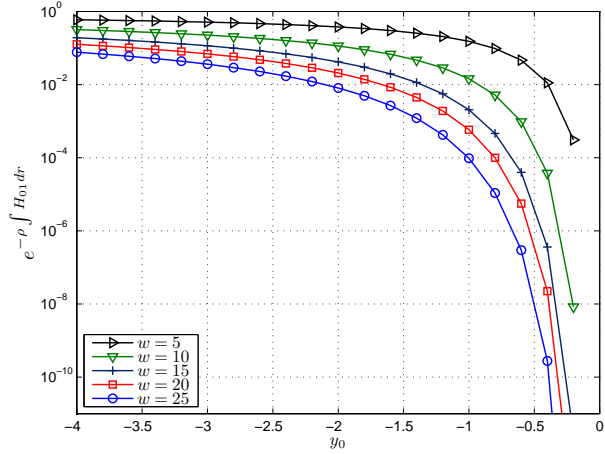}
	\caption{Effect of $y_0$ on the external node isolation probability, $\rho =0.1$, $w=20$, $L=100$, $\epsilon = 0.3$, $\alpha = 0.6$, $K =4$, $C=6$.}
	\label{fig:outage_escape_y0}
\end{figure}

Furthermore, we are interested in investigating the effects of each of the system parameters on the exterior node isolation probability.  We first analyze the effects of $y_0$ on this value, assuming $\alpha =0.6$, $\rho=0.1$, $\beta =10^{-3}$ and $K =4$ for the Rician fading channel.  Results are shown in Fig. \ref{fig:outage_escape_y0}.  As expected, a smaller value of $|y_0|$ leads to a decrease in $e^{-\rho\int H_{01}d r_1}$.  As node $0$ moves closer to the gap, the value of $\theta$ increases, resulting in a larger area being covered with successive numbers of reflections from the boundaries.  This in turn increases the probability of the external node being connected to one of the internal nodes.  Analysis of the plot also indicates that as the width of the system is increased (at a fixed density of $ \rho=0.1$), the probability that node $0$ does not connect to any other nodes decreases.  Such an observation is attributed to a larger LOS region for systems with larger values of $w$.  Similarly, the areas covered by successive reflections are increased as well.  As such, there is a higher chance that the external node can connect to another node on the interior as the signal travels a shorter path and undergoes a lower number of reflections.  On that account, the signal attenuation is lower.  

\begin{figure}[t]
	\centering
		\includegraphics[width=9cm]{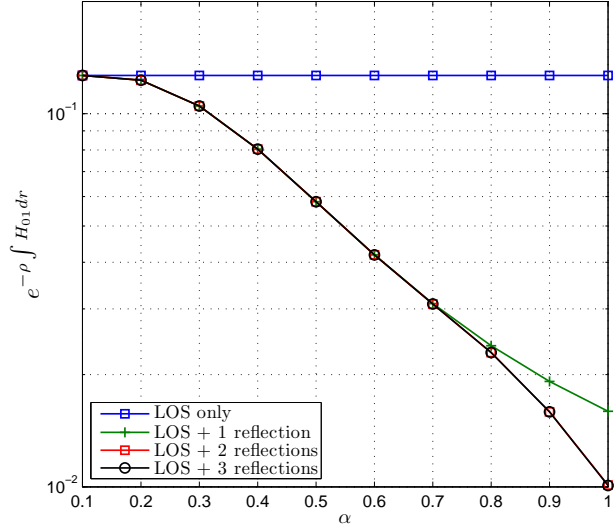}
	\caption{Effects of the number of reflection on the probability of node 0 not connecting to an internal node, $w =15$, $y_0=-2$, $L=100$, $\beta = 10^{-3}$, $\rho =0.1$, $\epsilon = 0.3$, $\theta = 0.0749$, $K=4$, $C=6$.}
	\label{fig:outage_escape_alpha_y0_2}
\end{figure}

The effects of individual reflections is investigated next.  To illustrate the benefits of considering reflections in the connectivity analysis, the effect of increasing numbers of reflections considered is illustrated in Fig. \ref{fig:outage_escape_alpha_y0_2}.  It can be seen from the figure that the probability of node $0$ not connecting to an internal node decreases as more reflections are considered, especially for large values of $\alpha$.  However, it can also be observed from the plot that the influence of three or more reflections is negligible on this outage probability.  This observation is due to the length scale relative to the system size, including the value of $\beta$ chosen for the simulations.  Choosing much smaller values for this parameter (for e.g., $10^{-5}$ or less) may lead to a higher connection probability as the number of reflections increases. Nevertheless, as previously stated, the focus of this contribution is on very high frequency communication systems.  Under such transmission parameters, a larger value of $\beta$ would arise.  Signal quality degradation as a function of distance covered is more pronounced.  Thus, even though the signal is not attenuated on hitting an obstacle ($\alpha=1$), the longer path of a reflected ray often implies that the received signal at a node far from the gap is often unrecoverable.

\begin{figure}[t]
	\centering
		\includegraphics[width=9cm]{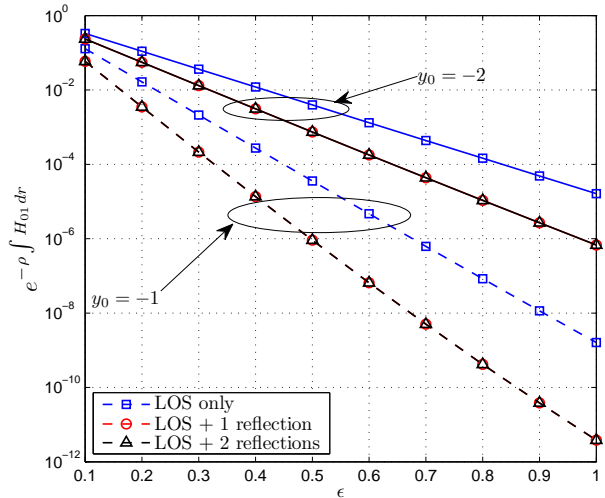}
	\caption{Effect of gap length on the probability of the node 0 not connecting to an internal node, $w=20$, $L=100$, $\beta = 10^{-3}$, $\rho =0.1$, $\alpha = 0.75$, $K=4$.}
	\label{fig:outage_escape_gapLength_y0_2} 
\end{figure}

Another system parameter expected to affect the connectivity of this network is the gap length, $\epsilon$.  The effect of this parameter is investigated in Fig. \ref{fig:outage_escape_gapLength_y0_2}.  The smaller the gap size (equivalent to a smaller value of $\theta$), the higher is the probability of node 0 not connecting to any other internal node.  Recognizing the logarithmic scale of the vertical axis in the plot, it can be deduced the exterior node isolation probability would decay exponentially with increasing size of the gap.  This can also be inferred from (\ref{eq:mass_H01}) where for $\theta\ll 1$, $\int H_{01} d\mathbf{r} = \mathcal{O}\left( \theta\right)$. 

\subsubsection{3-D Scenario}

\begin{figure}[t]
	\centering
		\includegraphics[width=9cm]{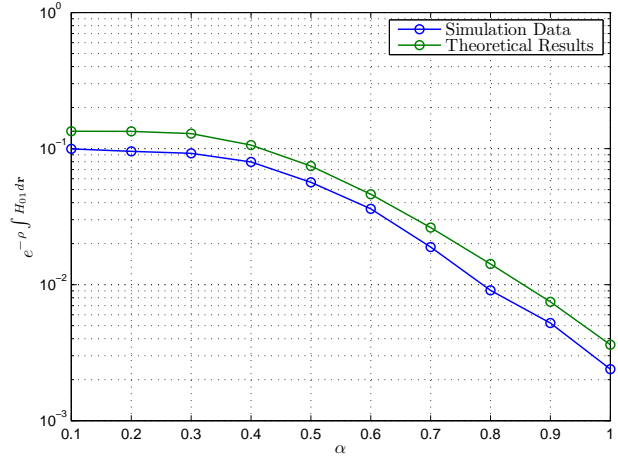}
	\caption{Comparison of outage probability based on theoretical results and simulation data for the 3-D case, with $w=20$, $L=100$, $\beta=10^{-3}$, $z_0 = -2$, gap radius $\epsilon =0.1$, $\rho=0.08$, $K=4$, $C=6$.}
	\label{fig:compare_simul_theory_escapa_out_3D}
\end{figure}

Using a similar procedure as in the preceding subsection, we analyze the suitability of the derived expression by comparing it with results from Monte Carlo simulations.  Results are shown in Fig. \ref{fig:compare_simul_theory_escapa_out_3D}.  In this plot, a circular gap of radius $0.1$ is assumed, $z_0$ is set to $-2$, $\beta = 10^{-3}$ and $K =4$.  Similar to the 2-D scenario, the plot indicates that the approximations used in the derivations of the connectivity mass are appropriate for the analysis of the system.

\begin{figure}[t]
	\centering
		\includegraphics[width=9cm]{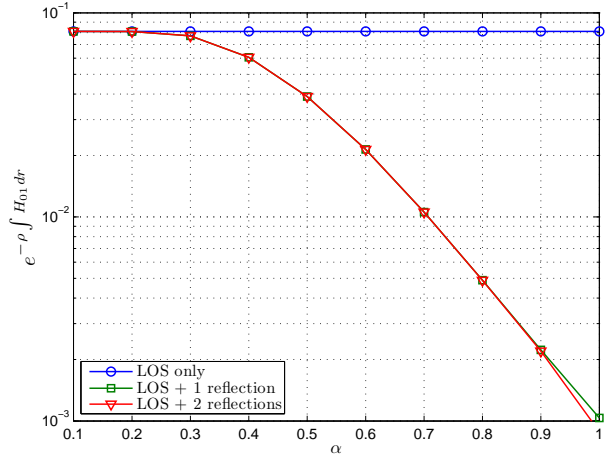}
	\caption{Analysis of the contribution of reflected signals on the outage probability for the 3-D case, $w=20$, $L=100$, $\beta=10^{-3}$, $z_0 = -2$, $\rho=0.1$, $\epsilon =0.1$, $K=4$, $C=6$.}
	\label{fig:outage_escape_alpha_z0_2_3D} 
\end{figure}

Following the same approach as for the 2-D scenario, we analyze the effects of the reflected rays on the outage probability of the system.  This effect is illustrated in Fig. \ref{fig:outage_escape_alpha_z0_2_3D} where the same parameter values as above are used.  As previously observed, the contributions of signals undergoing more than 2 reflections on the connectivity is negligible given the system parameters (c.f., Fig. \ref{fig:outage_escape_alpha_y0_2}).

\begin{figure}[t]
	\centering
		\includegraphics[width=9cm]{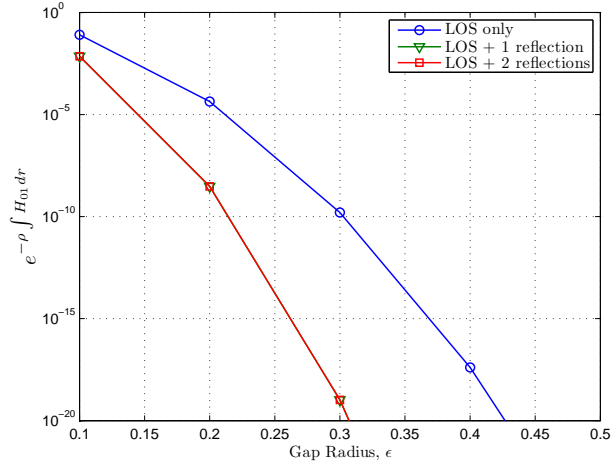}
	\caption{Effect of the size of the gap on the outage probability for the 3-D case, $w=20$, $L=100$, $\beta=10^{-3}$, $\alpha=0.75$, $z_0 = -2$, $\rho=0.1$, $K=4$, $C=6$.}
	\label{fig:outage_escape_gap_3D} 
\end{figure}

\begin{figure}[t]
	\centering
		\includegraphics[width=9cm]{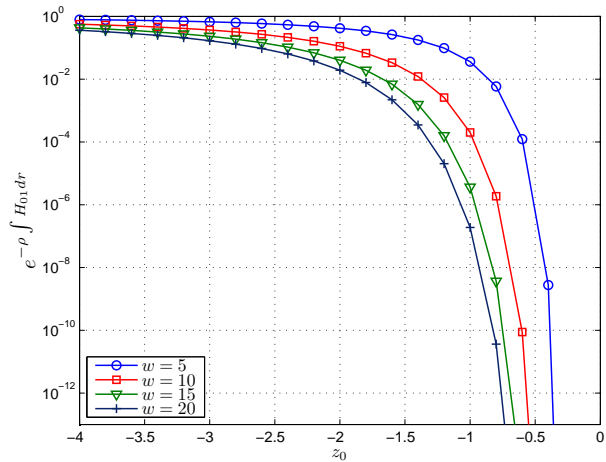}
	\caption{Effect of the distance of the external node to the gap, $z_0$, on the external node isolation probability in the 3-D system, $L=100$, $\beta=10^{-3}$, $\alpha=0.75$, $\rho=0.1$, $K=4$, $C=6$.}
	\label{fig:outage_escape_z0_3D}
\end{figure}

On the other hand, the effect of the radius of the gap in illustrated in Fig. \ref{fig:outage_escape_gap_3D}, while the relation between the outage probability and the distance of the external node from the gap is shown in Fig. \ref{fig:outage_escape_z0_3D}.  Similar to the 2-D system, an increase in the value of $w$ (corresponding to the height of the system) or increasing size of the gap results in a lower outage probability.  The reasons for such behavior is as previously stated, namely, a larger LOS region is obtained.  However, the rate of decrease of the outage probability in the 3-D case differs from the simple 2-D case.  Such an observation can be credited to the way in which the areas (volumes) of the spaces covered by signals undergoing reflections with the boundaries behave in the two systems.  To illustrate this point, let us consider the ratio of areas (volumes) covered by the LOS signals and signals that are reflected once from the boundaries.  For the 2-D case, we refer to the area covered by the LOS signals by $a_0$ while the area covered after one reflection is referred to as $a_1$.  Similarly, we use $v_0$ and $v_1$ to represent to respective volume of space in the 3-D system.  It can easily be shown that 
\begin{equation}
\frac{a_1}{a_0} = \frac{3w +2|y_0|}{w+2|y_0|} = \frac{3+2\frac{|y_0|}{w}}{1+2\frac{|y_0|}{w}} = 1+ \frac{2}{1+2\frac{|y_0|}{w}}
\end{equation}
while
\begin{eqnarray}
\frac{v_1}{v_0} &=& 1 + \frac{|z_0|^3 - 2(w+|z_0|)^3 + (2w+|z_0|)^3}{(w+|z_0|)^3-|z_0|^3} \nonumber\\
&=& 1 + \frac{6\left(1+\frac{|z_0|}{w} \right)}{1 + 3\frac{|z_0|}{w} + 3\left(\frac{|z_0|}{w} \right)^2}
\end{eqnarray}
Clearly, the rates at which the connection probabilities change with respect to $w$ while considering reflections are different in the two investigated scenarios.  In particular, in the case of the external node being close to the boundary, i.e., $\lim_{|y_0| \rightarrow 0}$ or $\lim_{|z_0| \rightarrow 0}$, the ratio in the above expressions limits to 2 for the $2-D$ case, while the ratio limits to $6$ in the $3-D$ case.

It should be noted that in the system model under investigation, there are numerous parameters that affect the full connectivity of the network.  Due to space restrictions, it is not possible to delve in each possible combination that would increase the connection probability.  The reader should however bear in mind that further analysis can be readily carried based on the expressions presented in this paper.

\section{Transport Problem}
\label{sec:transport}
In this section, we study the connectivity in the second scenario, namely when both transmitting and receiving nodes are located outside the main system.  This study is comparable to the work done in \cite{Dettmann2011} where the transport problem is investigated for the stadium billiard.  Such a scenario can have a number of practical applications. For instance, it may model the situation where nodes are located in different rooms next to small openings.  While it is possible to consider the case where multiple nodes are located within the main system and derive the full connection probability as for the escape problem, we restrict our analysis to the case where nodes are only located outside the system.  In doing so, we avoid the repetition of the results presented in the preceding section.  The derivation of the full connection probability of all nodes within the boundaries, and those external to the system close to the two gaps is a simple extension of the escape problem presented above. 

We investigate two cases for this problem.  In the first case, it is assumed that the gaps are on opposite sides of the boundaries, while in the second case, the gaps are assumed to be on the same side.  In either case, the consideration of reflected rays has a considerable impact on the connectivity of the network.   

\subsection{System Geometry}
\subsubsection{Case 1: Gaps on Opposite Sides}
\begin{figure}[t]
	\centering
		\includegraphics[width=9cm]{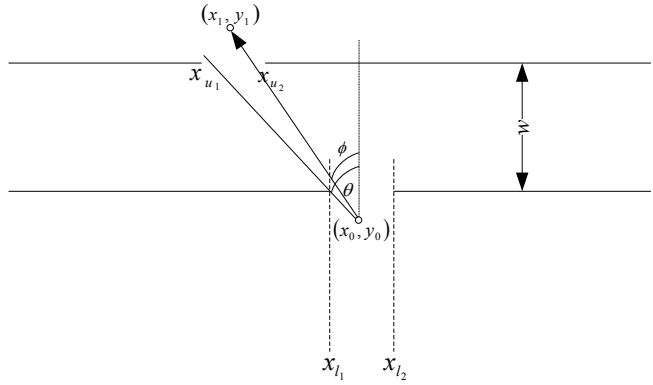}
	\caption{System model for the transport problem; case 1, where the gaps are on opposite sides of the boundaries.}
	\label{fig:transport_prob_case1}
\end{figure}

Consider the system model shown in Fig. \ref{fig:transport_prob_case1}.  Let the two nodes located at positions $(x_0,y_0)$ and $(x_1,y_1)$ represent a transmitter and receiver node respectively.  Assume that the positions of the gaps are fixed with the corresponding coordinates on the lower boundary of $x_{l_1}$ and $x_{l_2}$, and $x_{u_1}$ and $x_{u_2}$ for those on the upper boundary.  

Depending on the locations of the gaps and the nodes, a LOS transmission may not be possible, in which case, the only way for the nodes to communicate would be through reflections from the system boundaries. Similar to the escape problem, we assume that $x_{l_1}\leq x_0 \leq x_{l_2}$ and $x_{u_1}\leq x_1 \leq x_{u_2}$ and let the maximum angle of a ray escaping the gap measured from the vertical axis be $\theta = \tan^{-1}\frac{x_0-x_{l_1}}{|y_0|}$.
 
Referring to the figure, it is obvious that a direct connection is not possible if $x_0 - (|y_0|+w)\tan\theta > x_{u_2}$.  Under such conditions, the only trajectories for the signal would be through an even number of collisions.  Assuming the signal undergoes a total of $c$ reflections, the range of angles for a signal escaping the lower gap and entering the upper gap is
\begin{eqnarray}
\phi_{\min}^{(c)} \left(\phi \right) &=& \tan^{-1}\left(\frac{x_0 -x_{u_2}}{(c+1)w+|y_0|} \right)\\
\phi_{\max}^{(c)} \left(\phi \right) &=& \min\left(\theta,\tan^{-1}\left(\frac{x_0 -x_{u_1}}{(c+1)w+|y_0|} \right) \right)
\end{eqnarray}
The corresponding set of distances traveled would then be
\begin{eqnarray}
r_{\min}^{(c)} &=& \frac{(c+1)w+|y_0|}{\cos\phi} \\
r_{\max}^{(c)} &=& \frac{x_0 - x_{u_1}}{\sin \phi}
\end{eqnarray}
It should be noted that in the above, it was assumed that the gaps are not exactly opposite each other. If this were the case, only direct rays between the nodes would be possible since no reflected rays would enter the upper boundaries after an even number of reflections.

\subsubsection{Case 2: Gaps on the Same Side}
\begin{figure}[t]
	\centering
		\includegraphics[width=9cm]{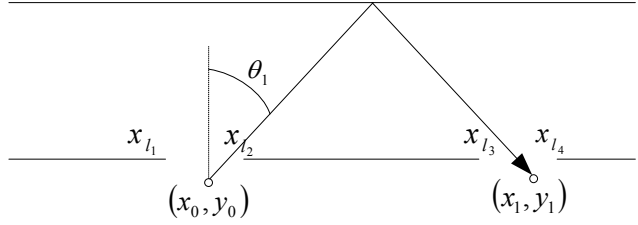}
	\caption{System model for the transport problem; case 2, where the gaps are on the same sides of the boundaries.}
	\label{fig:transport_prob_case2}
\end{figure}

In the second scenario, it is assumed that the gaps are on the same side of the boundaries. Thus, connection between nodes outside each gap is only possible after an odd number of reflections as shown in Fig. \ref{fig:transport_prob_case2}. Let the positions of the gap be $x_{l_1}$, $x_{l_2}$, $x_{l_3}$ and $x_{l_4}$.
Let the positions of the transmitter and receiver nodes be $(x_0,y_0)$ and $(x_1,y_1)$ respectively.  For a given transmitter node location, the maximum escape angle from the gap, $\theta$, is given by:
\begin{equation}
\theta = \tan^{-1} \left(\frac{x_{l_2} - x_0}{|y_0|} \right)
\end{equation}
The minimum escape angle which would allow more than one reflection to occur with the boundaries is:
\begin{equation}
\phi_{\min} = \tan^{-1} \left(\frac{x_{l_2} - x_0}{2w+|y_0|} \right)
\end{equation}
Given such a geometry, the connection between the two nodes would be possible only after an odd number of reflections.  The angle $\phi^{(c)}$, measured from the vertical axis, for the transmitter and receiver node to connect after $c$ reflections is given by
\begin{equation}
\phi^{(c)} = \tan^{-1} \left(\frac{x_1-x_0}{(c+1)w +|y_0| + |y_1|} \right)
\end{equation}
where $c$ is odd.  The minimum number of reflections such that connection would be possible satisfies the relationship that $\phi_{\min} \leq \phi^{(c)} \leq \theta$.  For each such path, the distance between the transmitter and receiver nodes is
\begin{equation}
r^{(c)} =  \left[(c+1)w+|y_1|+|y_0| \right]\sec\phi^{(c)}
\end{equation}
 
The same approach as in the previous section can be followed to determine the region covered in the receiver gap for a given transmitter node position and a given number of reflections.  In this case, the minimum and maximum escape angles of a signal from the transmitter gap are given by:
\begin{eqnarray}
\phi_{\min}^{(c)} &=& \tan^{-1}\left(\frac{x_{l_3}-x_0}{(c+1)w+|y_0|} \right) \\
\phi_{\max}^{(c)} &=& \min\left(\theta, \tan^{-1}\left( \frac{x_{l_4}-x_0}{(c+1)w+|y_0|}\right) \right)
\end{eqnarray}
while the range of distance for a given angle $\phi$ is
\begin{eqnarray}
r_{\min}^{(c)}\left(\phi \right) &=& \frac{x_{l_3} - x_0}{\sin \phi}  = \frac{(c+1)w+|y_0|}{\cos \phi}\\
r_{\max}^{(c)}\left(\phi \right) &=& \frac{x_{l_4} - x_0}{\sin \phi}
\end{eqnarray}
where $x_{l_3}$, $x_{l_4}$ and $x_0$ are assumed fixed.  

\subsection{Analysis of Connection Probability}
Let a single node be located close to each gap.  Given the geometries considered, it can be assumed that a Rician fading channel is present, where the signal undergoing the shortest path and minimum number of reflections is the strongest (equivalent to a LOS) signal.  Similar to the derivations in the previous sections, the connection probability between a transmitter and receiver node given a minimum acceptable data rate of $R_0$ is $H_{01} \approx \exp\left(e^{-\nu} b^{\mu}\right)$.  The connection probability averaged over all configurations between the pair of transmitter and receiver nodes can be expressed as:
\begin{equation}
\frac{1}{V_0 V_1}\int_{\mathbf{r}_0}\int_{\mathbf{r}_1} H_{01} d\mathbf{r}_1d\mathbf{r}_0
\end{equation}
where $V_0$ and $V_1$ are the areas of the regions where the transmitter and receiver nodes can reside.  It is straightforward to extend the above derivations to determine the full connectivity if nodes are present within the system boundaries.  In that case, each gap can be analyzed using the same approach as the escape problem as derived previously. 

\subsubsection{Case 1}
Assuming that the gap locations and sizes are fixed, for a given node $0$ position, the integral of $H_{01}$ yields 
\begin{eqnarray}
\int_{\mathbf{r}_1} H_{01} d\mathbf{r}_1 &=& \mathop{\sum_{c=c_{\min}}^C}_{c \text{ is even} }\int_{\phi_{\min}^{(c)}}^{\phi_{\max}^{(c)}} \int_{r_{\min,\phi}^{(c)}}^{r_{\max, \phi}^{(c)}} r H_{01} dr d\phi \nonumber\\
&=& \frac{1}{\mu} \mathop{\sum_{c=c_{\min}}^C}_{c \text{ is even} } \int_{\phi_{\min}^{(c)}}^{\phi_{\max}^{(c)}} \lambda_c^{-\frac{2}{\mu}} \left[\gamma\left(\frac{2}{\mu},\left(r^{(c)}_{\max}\right)^{\mu} \lambda_c \right) - \gamma\left(\frac{2}{\mu},\left(r^{(c)}_{\min}\right)^{\mu} \lambda_c \right) \right] d\phi \nonumber\\
&=& \frac{1}{\mu} \mathop{\sum_{c=c_{\min}}^C}_{c \text{ is even} } \int_{\phi_{\min}^{(c)}}^{\phi_{\max}^{(c)}} \lambda_c^{-\frac{2}{\mu}} \left[\gamma\left(\frac{2}{\mu},\left(\frac{x_0 - x_{u_1}}{\sin \phi}\right)^{\mu} \lambda_c \right) - \gamma\left(\frac{2}{\mu},\left( \frac{(c+1)w+|y_0|}{\cos\phi}\right)^{\mu} \lambda_c \right) \right] d\phi
\label{eq:int_transport_case1}
\end{eqnarray}
where $c_{\min}$ is the minimum number of reflections that would allow connection between the two nodes.  To perform the integration with respect to $\phi$, a series expansion can be used again.  However, in contrast to the escape problem where it was reasonable to assume that $\phi$ would be close to 0, such an assumption may not always be valid in this scenario.  At the expense of less concise expressions, the expansion is performed about $\tilde{\phi}_c = \frac{\phi_{\min}^{(c)} + \phi_{\max}^{(c)}}{2}$ and given in (\ref{eq:gamma_inc_transport}).
\begin{eqnarray}
\gamma\left(\frac{2}{\mu},\left(\frac{x_0 - x_{u_1}}{\sin \phi}\right)^{\mu} \lambda_c\right) &\approx& \gamma\left(\frac{2}{\mu},\csc^{\mu}(\tilde{\phi}_c)(x_0 - x_{u_1})^{\mu}\lambda_c \right) - A_t (\phi-\tilde{\phi}_c) \cot(\tilde{\phi}_c) \nonumber\\
&& + \frac{A_t B_t}{2}(\phi-\tilde{\phi}_c)^2 + \mathcal{O}(\phi-\tilde{\phi}_c)^3 \nonumber\\
\gamma\left(\frac{2}{\mu},\lambda_c \left((c+1)w+|y_0| \right)\sec^{\mu}(\phi) \right) &\approx& \gamma\left(\frac{2}{\mu},\lambda_c\left((c+1)w+|y_0| \right)^{\mu}\sec^{\mu}(\tilde{\phi}_c) \right) + D_t \tan(\tilde{\phi}_c)(\phi-\tilde{\phi}_c) \nonumber\\
&& - \frac{D_tE_t}{2}(\phi-\tilde{\phi}_c)^2
\label{eq:gamma_inc_transport}
\end{eqnarray}
where 
\begin{eqnarray}
A_t &=& e^{-\lambda_c (x_0-x_{u_1})^{\mu}\csc^{\mu}(\tilde{\phi}_c)} \mu\lambda_c^{\frac{2}{\mu}} (x_0-x_{u_1})^2\csc^2(\tilde{\phi}_c)\nonumber\\
B_t &=& \csc^2(\tilde{\phi}_c) \left[2+\cos(2\tilde{\phi}_c) -\mu\lambda_c\cos^2(\tilde{\phi}_c)(x_0-x_{u_1})^{\mu} \csc^{\mu}(\tilde{\phi}_c) \right]\nonumber\\
D_t &=& e^{-\lambda_c\left( (c+1)w+|y_0|\right)^{\mu}\sec^{\mu}(\tilde{\phi}_c)} \mu\lambda_c^{\frac{2}{\mu}}\left((c+1)w+|y_0| \right)^2\sec^2(\tilde{\phi}_c)\nonumber\\
E_t &=& \sec^2(\tilde{\phi}_c) \left[-2+\cos(2\tilde{\phi}_c) + \mu\lambda_c \left((c+1)w+|y_0| \right)^{\mu}\sec^{\mu}(\tilde{\phi}_c) \sin^2(\tilde{\phi}_c) \right] \nonumber
\end{eqnarray}
Substituting the above approximations in (\ref{eq:int_transport_case1}) results in: 
\begin{align}
\label{eq:int_H01_trans}
\int_{\mathbf{r}_1} & H_{01} d\mathbf{r}_1 = \frac{1}{\mu} \mathop{\sum_{c=c_{\min}}^C}_{c \text{ is even} }\left(\phi_{\max}^{(c)} - \phi_{\min}^{(c)} \right)\left(\gamma\left(\frac{2}{\mu},\csc^{\mu}(\tilde{\phi}_c)(x_0-x_{u_1})^{\mu}\lambda_c \right) - \gamma\left(\frac{2}{\mu},\lambda_c \left((c+1)w+|y_0| \right)^{\mu}\sec^{\mu}(\tilde{\phi}_c) \right) \right) \nonumber\\
& - \left(A_t\cot(\tilde{\phi}_c) + D_t\tan(\tilde{\phi}_c)\right)\left(\frac{(\phi_{\max}^{(c)})^2}{2} - \phi_{\max}^{(c)}\tilde{\phi}_c - \frac{(\phi_{\min}^{(c)})}{2} + \phi_{\min}^{(c)}\tilde{\phi}_c \right) \nonumber\\
&+ \frac{1}{6} \left( A_t B_t+D_t E_t\right) \left[ \left( \phi_{\max}^{(c)} -\tilde{\phi}_c \right)^3 - \left( \phi_{\min}^{(c)} -\tilde{\phi}_c \right)^3  \right] + \mathcal{O}(\phi_{\max}^{(c)} - \phi_{\min}^{(c)})^4 \nonumber\\
&\approx \frac{1}{\mu} \mathop{\sum_{c=c_{\min}}^C}_{c \text{ is even} }\left(\phi_{\max}^{(c)} - \phi_{\min}^{(c)} \right)\left(\gamma\left(\frac{2}{\mu},\csc^{\mu}(\tilde{\phi}_c)(x_0-x_{u_1})^{\mu}\lambda_c \right) - \gamma\left(\frac{2}{\mu},\lambda_c \left((c+1)w+|y_0| \right)^{\mu}\sec^{\mu}(\tilde{\phi}_c) \right) \right) 
\end{align}

\begin{figure}[t]
	\centering
		\includegraphics[width=9cm]{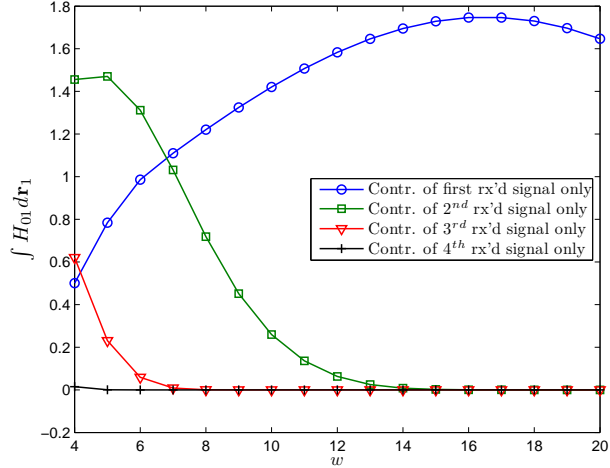}
	\caption{Effect of different numbers of reflections on the connectivity mass for the transport problem, $L=100$, $y_0=-2$, $\epsilon = 0.3$, $\beta=10^{-3}$, $\alpha=0.85$, $K=4$.}
	\label{fig:contr_reflec_trans1_alpha_085}
\end{figure}

The novelty in this approach to analyzing wireless connectivity is the consideration of signal reflections from the boundaries.  To illustrate this effect, the contribution of each received signal to $\int H_{01} d\mathbf{r}_1$ after undergoing a given number of reflections is shown in Fig. \ref{fig:contr_reflec_trans1_alpha_085} for different values of $w$.  For these plots, a Rician fading model with $K=4$ was assumed, with $\beta = 10^{-3}$, $\alpha = 0.85$, $|y_0| =2$, gap lengths of $0.3$, $x_{l_1} = 15$, $x_{u_1} =14.5 $, while the corresponding value of $\theta = 0.0749$.  Under these simulation parameters, the minimum number of reflections for every value of $w$ considered is $0$.  Thus, the reference to the first received signal in the plot corresponds to the LOS signal, the second received signal is the one that undergoes reflections and so on.  As expected, the effect of signals undergoing higher number of reflections is less significant.

\begin{figure}[t]
	\centering
		\includegraphics[width=9cm]{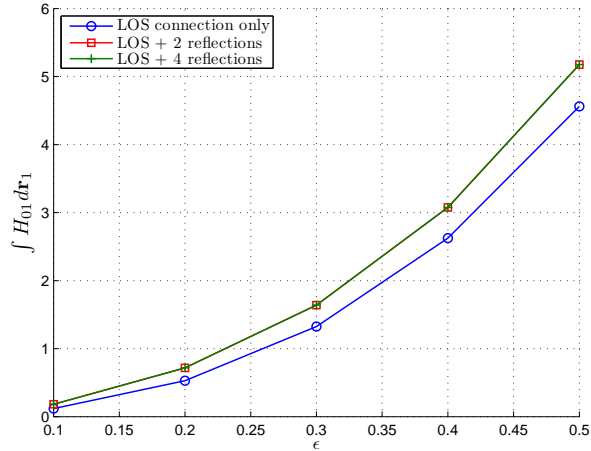}
	\caption{Effect of gap length on the connectivity mass for the transport problem, $w=10$, $L=100$ $y_0=-2$, $\beta=10^{-3}$, $\alpha=0.85$, $K=4$.}
	\label{fig:contr_reflec_trans1_alpha_085_gapLength}
\end{figure}

We next investigate the effect of the size of the gap on the connectivity mass.  This is illustrated in Fig. \ref{fig:contr_reflec_trans1_alpha_085_gapLength}.  Similar to previous observations, it can be seen that the larger the gap, the larger is the value of the integral $\int H_{01}d\mathbf{r}_1$.  Consequently, the outage probability decreases with increasing gap length.  Such a behavior is intuitive since a larger gap length is synonymous of a larger LOS region.  

\begin{figure}[t]
	\centering
		\includegraphics[width=8cm]{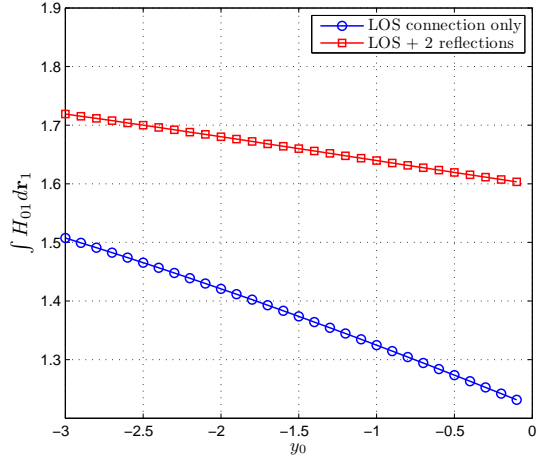}
	\caption{Effect of $y_0$ on the connectivity mass for the transport problem, $\epsilon =0.3$, $w=10$, $L=100$ $\beta =10^{-3}$, $\alpha=0.85$, $K=4$.}
	\label{fig:contr_reflec_trans1_alpha_085_y0}
\end{figure}

In Fig. \ref{fig:contr_reflec_trans1_alpha_085_y0}, we present the effect of the distance $y_0$ on the connectivity mass of the system.  For these sets of simulations, we assume that $\alpha = 0.85$.  From the plot, it can be observed that as node 0 moves closer to the boundary of the system, the connectivity mass decreases, which implies a larger outage probability.  Such an observation is different to what was observed for the escape problem.  This behavior can however be easily explained by considering the geometry of the system.  Given the positioning of the gaps (same as in the above simulations), moving node 0 closer to the gap actually leads to smaller coverage area for the LOS region which leads to a smaller connectivity mass.  On the other hand, moving this node closer to the boundary also leads to a wider area being covered by the reflected paths that can enter the gap on the upper boundary.  This explains the different rates at which the two curves decrease in the figure.  

\section{Conclusion}
\label{sec:conclusion}
In this paper, the connectivity of networks in confined geometries has been investigated wherein signal reflections from the boundaries have been explicitly considered.  In particular, the full connection probability of a network where at least one node is located outside the main geometry has been investigated.  This paper has provided an in depth mathematical analysis of how the connectivity of such networks is affected by various system parameters, such as the physical dimensions of the geometry, the size of the opening on the boundary and the reflectivity properties of the boundaries.  Although the connectivity was initially studied for a simple 2-D system, it was shown that the analysis can be easily extended to more practical 3-D systems.  It is believed that this contribution will provide the framework for further analysis of the connection probability of networks residing in convex as well as non-convex  geometries.

\section*{Acknowledgement}

The authors would like thank Toshiba Telecommunications Research Laboratory
and the EPSRC (grant EP/H500316/1) for their continued support.

\appendix
\label{sec:appA}
\textbf{Derivation of the Second Average Term in (\ref{eq:full_pc})}\\
We present the method of evaluating the second average term in (\ref{eq:full_pc}) in this section.  As previously stated, this average term would be negligible under the system configuration considered in this paper.  This term refers to the probability that one internal node is not connected to the remaining nodes and involves the evaluation of the integral $\int H_{1N} d\mathbf{r}_1$, where $H_{1N}$ is the pair-connectedness function between nodes 1 and $N$, both of which are internal to the system.  
Since the space enclosed by the boundaries is convex in our model, a LOS path or direct connection is possible between any pair of nodes\footnote{Throughout this paper, we have assumed that any node does not shadow the signals from other nodes.}.  On the account that reflected signals would typically be weaker than the direct or LOS signal, reflections from the boundaries are not explicitly considered in this part of the analysis.  However, a Rician fading model is still assumed.  In that case, $H_{1N}= Q_1\left(\sqrt{2K}, r\sqrt{2(K+1)\beta} \right)$, which leads to
\begin{eqnarray}
\int H_{1N}d\mathbf{r}_1 &=& \int_0^w \int_0^L \exp\left(\nu r^{\mu} (2(K+1)\beta)^{\frac{\mu}{2}} \right) dx_1 dy_1 \nonumber\\
&=& \int_0^w \int_0^L \exp\left(\hat{\lambda} \left((x_1 - x_N)^2 +(y_1 -y_N)^2 \right)^{\frac{\mu}{2}} \right)dx_1 dy_1 
\end{eqnarray}
where $\hat{\lambda} = \nu (2(K+1)\beta)^{\mu/2}$.  It is realized that evaluating the above integrals can be very challenging, even while using the approximation of the Marcum $Q-$function as performed earlier.  While the connectivity mass under the assumption of Rayleigh fading for a square has been derived in \cite{Coon2012}, it is not possible to directly extend the work to the Rician fading case.  Given the difficulty in evaluating the integral of the above exponential term, another approximation to the Marcum $Q$-function is made, namely, $Q(a,b)\approx \exp\left(-e^{\nu_2} b^{2}\right)$.  Compared to the previous approximation, the exponent on the variable $b$ has been changed from $\mu$ to $2$.  Similar to the method used in \cite{Bocus2013}, the derivation of the parameter $\nu_2$ for a given value of $K$ can be obtained by minimizing the sum of square errors (SSE) between the actual and approximate function.  With this approximation,
\begin{eqnarray}
\int H_{1N} d\mathbf{r}_1 &=& \int_0^w \int_0^L e^{-\hat{\lambda}(x_1-x_N)^2 }e^{-\hat{\lambda}(y_1-y_N)^2 } dx_1 dy_1 \nonumber\\
&=& \frac{\pi}{4\hat{\lambda}} f(x_N,L) f(y_N,w)
\end{eqnarray}
where $f(x,l) = \left(\text{erf}\left((l-x)\sqrt{\hat{\lambda}}\right)  + \text{erf} \left( x\sqrt{\hat{\lambda}} \right) \right)$.  The first integral term in (\ref{eq:H_1N_express}) then becomes
\begin{align}
\rho & \int e^{-\rho \int H_{1N}d\mathbf{r}_1} d\mathbf{r}_N = \rho \int_0^w \int_0^L \exp\left(-\rho\frac{\pi}{4\hat{\lambda}} f(x_N,L) f(y_N,w) \right) d x_N dy_N 
\end{align}
To evaluate the above integral, the series expansion of $f(x_N,L)$ and $f(y_N,w)$ about $L/2$ and $w/2$ respectively can be taken, which corresponds to the points at which the functions are maximum.    
\begin{align}
f(x_N,L) &\approx  2\text{erf}\left(\sqrt{\hat{\lambda}}\frac{L}{2} \right) -   \left(x_N-\frac{L}{2}\right)^2 \tau_1 + \mathcal{O}\left(x_N-\frac{L}{2} \right)^3 \nonumber\\
f(y_N,w) &\approx  2\text{erf}\left(\sqrt{\hat{\lambda}}\frac{w}{2} \right) -   \left(y_N-\frac{w}{2}\right)^2 \tau_2 + \mathcal{O}\left(y_N-\frac{w}{2} \right)^3 
\end{align}
where $\tau_1 = \frac{2}{\sqrt{\pi}} L\hat{\lambda}^{\frac{3}{2}}e^{-\frac{\hat{\lambda} L^2}{4}} $ and $\tau_2 = \frac{2}{\sqrt{\pi}} w\hat{\lambda}^{\frac{3}{2}}e^{-\frac{\lambda w^2}{4}} $.  Neglecting the terms with the product of variables $x_N$ and $y_N$ in the expansion of $f(x_N,L) f(y_N,w) $ leads to
\begin{eqnarray}
\int_0^w \int_0^L e^{-\rho \int H_{1N}d\mathbf{r}_1}dx_N dy_N &\approx& \int_0^w \int_0^L e^{-\rho\frac{\pi}{4\hat{\lambda}} \left[4 \text{erf}\left(\sqrt{\hat{\lambda}}\frac{L}{2} \right) \text{erf}\left(\sqrt{\hat{\lambda}}\frac{w}{2}\right) - 2 \tau_2\text{erf}\left(\sqrt{\hat{\lambda}} \frac{L}{2} \right)(y_N-\frac{w}{2})^2 - 2 \tau_1\text{erf}\left(\sqrt{\hat{\lambda}} \frac{w}{2} \right)(x_N-\frac{L}{2})^2  \right] } \nonumber\\
&=&e^{-\rho\frac{\pi}{\hat{\lambda}} \text{erf}\left(\sqrt{\hat{\lambda}}\frac{L}{2} \right) \text{erf}\left(\sqrt{\hat{\lambda}}\frac{w}{2}\right)} \int_0^w \int_0^L e^{\sigma_1\left(y_N-\frac{w}{2}\right)^2} e^{\sigma_2\left(x_N-\frac{L}{2}\right)^2} dx_N dy_N \nonumber\\
&=&e^{-\rho\frac{\pi}{\hat{\lambda}} \text{erf}\left(\sqrt{\hat{\lambda}}\frac{L}{2} \right) \text{erf}\left(\sqrt{\hat{\lambda}}\frac{w}{2}\right)} \int_{-\frac{w}{2}\sqrt{\sigma_1}}^{\frac{w}{2}\sqrt{\sigma_1}} \int_{-\frac{L}{2}\sqrt{\sigma_2}}^{\frac{L}{2}\sqrt{\sigma_2}} \frac{1}{\sqrt{\sigma_1 \sigma_2}} e^{z_1^2 + z_2^2} dz_1 dz_2 \nonumber \\
&=&\frac{2}{\sqrt{\sigma_1 \sigma_2}} e^{-\rho\frac{\pi}{\hat{\lambda}} \text{erf}\left(\sqrt{\hat{\lambda}}\frac{L}{2} \right) \text{erf}\left(\sqrt{\hat{\lambda}}\frac{w}{2}\right)} \Bigg[\int_0^{\vartheta} \left(e^{\frac{\sigma_2 L^2}{4} \sec^2 \phi} -1\right) d\phi +\nonumber\\
&& \int_{\vartheta}^{\frac{\pi}{2}} \left(e^{\frac{\sigma_1 w^2}{4} \csc^2 \phi} -1\right) d\phi \Bigg] 
\label{eq:H1n_part1}
\end{eqnarray}
where $\sigma_1 = \rho\tau_1\frac{\pi}{2\hat{\lambda}} \text{erf}\left(\sqrt{\hat{\lambda}}\frac{L}{2} \right)$, $\sigma_2 = \rho\tau_2\frac{\pi}{2\hat{\lambda}} \text{erf}\left(\sqrt{\hat{\lambda}}\frac{w}{2} \right)$ and $\vartheta = \tan^{-1}\left(\frac{w\sqrt{\sigma_1}}{L\sqrt{\sigma_2}} \right)$.  The last expression in (\ref{eq:H1n_part1}) is obtained by changing the integration to the polar coordinate system.  Using
\begin{eqnarray}
e^{a \sec^2{\phi}} &\approx& e^a \left(1+ a\phi^2\right) + \mathcal{O}(\phi^3) \\
e^{b \csc^2{\phi}} &\approx& e^{b\csc^2\vartheta} - (\phi -\vartheta)A_2 + (\phi -\vartheta)^2 B_2
\end{eqnarray}
where the series expansion is taken about $\phi =0$ and $\phi=\vartheta$ respectively, $A_2 = 2\left(b e^{b\csc^2\vartheta}\cot\vartheta \csc^2\vartheta \right)$ and $B_2 = b e^{b\csc^2\vartheta}\csc^2\vartheta\left(1+3\cot^2\vartheta + 2b\cot^2\vartheta\csc^2\vartheta \right)$, (\ref{eq:H1n_part1}) approximates to (\ref{eq:H1n_part1_approx}).
\begin{equation}
\frac{2}{\sqrt{\sigma_1 \sigma_2}} e^{-\rho\frac{\pi}{\hat{\lambda}} \text{erf}\left(\sqrt{\hat{\lambda}}\frac{L}{2} \right) \text{erf}\left(\sqrt{\hat{\lambda}}\frac{w}{2}\right)} \left[\begin{array}{l}\vartheta \left(e^{\frac{\sigma_2 L^2}{4}}-1 \right) + \frac{\sigma_2 L^2}{12}e^{\frac{\sigma_2 L^2}{4}}\vartheta^3 + \left(\frac{\pi}{2}-\vartheta \right)\left(e^{\frac{\sigma_1 w^2}{4} \csc^2\vartheta} + A_2\vartheta -1 \right) \\
-\frac{A_2}{2}\left(\frac{\pi^2}{4} -\vartheta^2 \right) + \frac{B_2}{3} \left(\frac{\pi}{2} - \vartheta \right)^3 \end{array}\right]
\label{eq:H1n_part1_approx}
\end{equation}

We now look at the last term in (\ref{eq:H_1N_express}).  This integral involves two terms, namely $H_{0N}$ and $H_{1N}$, which corresponds to the probability of the external node connecting to an internal node and the probability of any pair of internal nodes being connected.  However, as we have identified above, reflections from the system boundaries are considered only for the $H_{0N}$ term.  
\begin{equation}
\rho  \int H_{0N}\left(1-\frac{1}{V}\int H_{1N}d\mathbf{r}_1 \right)^{N-1} d\mathbf{r}_N = \rho \int \int H_{0N} e^{-\rho\int H_{1N} d\mathbf{r}_1} dx_N dy_N
\label{eq:H0n_H1n_integral}
\end{equation}
while the term $e^{-\rho\int H_{1N} d\mathbf{r}_1}$ and its integral have been defined above.  Consequently, the integral of (\ref{eq:H0n_H1n_integral}) comprises a sum of integrals over different $\mathcal{D}_c$, for $c=0,1,\cdots, C$ as
\begin{equation}
\rho \int  H_{0N}\left(1-\frac{1}{V}\int H_{1N}d\mathbf{r}_1 \right)^{N-1} d\mathbf{r}_N = \rho \sum_{c=0}^C \int_{\mathcal{D}_c} e^{-\bar{\lambda}_c \tilde{r}} e^{-\rho \int H_{1N}d\mathbf{r}_1} d\mathbf{r}_N
\end{equation}
where $\tilde{r}$ is the effective distance between node 0 and a corresponding internal node and $\bar{\lambda}_c = 2(K+1)\beta\alpha^{-c} \nu$.  Based on our previous observation, we consider a maximum of $2$ reflections and assume that node 0 lies in the middle of the gap.  Assuming that $\mathcal{D}_c$ is bounded in the $x-$direction by $x=l_b^{(c)}$ and $x=u_b^{(c)}$ (c.f. Section II), the required integral becomes 
\begin{eqnarray}
&&\rho \int H_{0N}\left(1-\frac{1}{V}\int H_{1N}d\mathbf{r}_1 \right)^{N-1} d\mathbf{r}_N =\nonumber\\
&& 2 e^{-\rho\frac{\pi}{\lambda} \text{erf}\left(\sqrt{\lambda}\frac{L}{2} \right) \text{erf}\left(\sqrt{\lambda}\frac{w}{2}\right)} \int_0^w \left[ \begin{array}{l}\int_{l_b^{(0)}}^{u_b^{(0)}} e^{-\bar{\lambda}_0\left((x_N-x_0)^2 + (y_N-y_0)^2 \right)} e^{\sigma_1\left(y_N -\frac{w}{2} \right)^2} e^{\sigma_2\left(x_N -\frac{L}{2} \right)^2} dx_N \\
+ \int_{l_b^{(1)}}^{u_b^{(1)}} e^{-\bar{\lambda}_1\left((x_N-x_0)^2 + (2w-y_N-y_0)^2 \right)} e^{\sigma_1\left(y_N -\frac{w}{2} \right)^2} e^{\sigma_2\left(x_N -\frac{L}{2} \right)^2} dx_N \\
+ \int_{l_b^{(2)}}^{u_b^{(2)}} e^{-\bar{\lambda}_2\left((x_N-x_0)^2 + (2w+y_N-y_0)^2 \right)} e^{\sigma_1\left(y_N -\frac{w}{2} \right)^2} e^{\sigma_2\left(x_N -\frac{L}{2} \right)^2} dx_N \end{array}\right] dy_N
\label{eq:prod_H0n_H1n}
\end{eqnarray}
The first term is the square bracket in (\ref{eq:prod_H0n_H1n}) can be expressed as in (\ref{eq:square_bracket_term})\footnote{Since we are not interested in average term in (\ref{eq:H_1N_express}) in this contribution and due to space restriction, we do not present the full simplification of expression herein.}.
\begin{align}
\int_0^w\int_{l_b^{(0)}}^{u_b^{(0)}} & e^{-\bar{\lambda}_0\left((x_N-x_0)^2 + (y_N-y_0)^2 \right)} e^{\sigma_1\left(y_N -\frac{w}{2} \right)^2} e^{\sigma_2\left(x_N -\frac{L}{2} \right)^2} dx_N dy_N \nonumber\\
&= e^{-q_x - q_y^{(0)}}\int_0^w \int_{l_b^{(0)}}^{u_b^{(0)}} e^{(\sigma_2-\bar{\lambda}_0)(x_N-p_x)^2  + (\sigma_1-\bar{\lambda}_0)(y_N-p_y^{(0)})^2 } dx_N dy_N \nonumber \\
&= \frac{e^{-q_x - q_y^{(0)}}}{\sqrt{(\sigma_1-\bar{\lambda}_0)(\sigma_2-\bar{\lambda}_0)}}  \int_{-\sqrt{\sigma_1-\bar{\lambda}_0} p_y^{(0)}}^{\sqrt{\sigma_1-\bar{\lambda}_0} (w-p_y^{(0)})} \int_{\sqrt{\sigma_2-\bar{\lambda}_0}(l_b^{(0)}-p_x)}^{\sqrt{\sigma_2-\bar{\lambda}_0}(u_b^{(0)}-p_x)} e^{z_1^2+z_2^2} dz_1 dz_2\nonumber\\
\label{eq:square_bracket_term}
\end{align}
where $p_x = \frac{\sigma_2 L-\bar{\lambda}_0 x_0}{(\sigma_2 - \bar{\lambda}_0)}$, $q_x = \sigma_2 \bar{\lambda}_0 \frac{(x_0 -L/2)^2}{(\sigma_2 - \lambda_0)}$, $p_y^{(0)} = \frac{\sigma_1 w-\bar{\lambda}_0 y_0}{(\sigma_1 - \bar{\lambda}_0)}$ and $q_y^{(0)} = \sigma_1 \bar{\lambda}_0 \frac{(y_0 -w/2)^2}{(\sigma_1 - \lambda_0)}$.

\bibliographystyle{IEEEtran}
\bibliography{IEEEabrv,library}

\end{document}